\documentclass[reprint,pra,superscriptaddress]{revtex4-1}

\usepackage[utf8]{inputenc}
\usepackage{booktabs}
\usepackage{mathtools}
\usepackage{amsmath, amssymb}
\usepackage{graphicx}
\usepackage{color}
\usepackage{array}
\usepackage[bookmarks=false]{hyperref}
\DeclarePairedDelimiter\bra{\langle}{\rvert}
\DeclarePairedDelimiter\ket{\lvert}{\rangle}
\DeclarePairedDelimiterX\braket[2]{\langle}{\rangle}{#1 \delimsize\vert #2}
\DeclarePairedDelimiter\mean{\langle}{\rangle}
\renewcommand*{\Re}{\mathrm{Re}}

\newcommand*{\up}{\uparrow}
\newcommand*{\down}{\downarrow}

\newcolumntype{M}[1]{>{\centering\arraybackslash}m{#1}}
\newcolumntype{N}{@{}m{0pt}@{}}

\begin{document}

\title{Unconventional quantum optics in topological waveguide QED}

\author{M.~Bello}
\affiliation{Instituto de Ciencia de Materiales de Madrid, CSIC, 28049, Spain}
\author{G.~Platero}
\affiliation{Instituto de Ciencia de Materiales de Madrid, CSIC, 28049, Spain}
\author{J.~I.~Cirac}
\affiliation{Max-Planck-Institut f\"ur Quantenoptik, Hans-Kopfermann-Strasse 1, 85748 Garching, Germany}
\author{A. Gonz\'alez-Tudela}
\affiliation{Max-Planck-Institut f\"ur Quantenoptik, Hans-Kopfermann-Strasse 1, 85748 Garching, Germany}
\affiliation{Instituto de F\'isica Fundamental IFF-CSIC, Calle Serrano 113b, Madrid 28006, Spain.}
\email{a.gonzalez.tudela@csic.es}
\date{\today}

\begin{abstract}
  The discovery of topological materials has challenged our understanding of
  condensed matter physics and led to novel phenomena. This has motivated recent
  developments to export topological concepts into photonics to make light
  behave in exotic ways. Here, we predict several unconventional quantum optical
  phenomena that occur when quantum emitters interact with a topological
  waveguide QED bath, namely, the photonic analogue of the Su-Schrieffer-Heeger
  model. When the emitters' frequency lies within the topological band-gap, a
  chiral bound state emerges, which is located at just one side (right or left)
  of the emitter. In the presence of several emitters, this bound state mediates
  topological, tunable interactions between them, that can give rise to exotic
  many-body phases such as double N\'eel ordered states. Furthermore, when the
  emitters' optical transition is resonant with the bands, we find
  unconventional scattering properties and different super/subradiant states
  depending on the band topology. Finally, we propose several implementations
  where these phenomena can be observed with state-of-the-art technology.   
\end{abstract}

\maketitle

\section{Introduction}

Even though the introduction of topology in condensed matter was originally
motivated to explain the integer Quantum Hall effect~\cite{klitzing80a}, its
implications were more far-reaching than expected. On a fundamental level,
topological order resulted in a large variety of new phenomena, as well as new
paradigms for classifying matter phases~\cite{wen17a}. On practical terms,
topological states can be harnessed to achieve more robust electronic devices or
fault-tolerant quantum computation~\cite{kitaev01a}. This
spectacular progress motivated the application of topological ideas to
photonics, for example, to engineer unconventional light behaviors. The starting
point of the field was the observation that topological bands also appear with
electromagnetic waves~\cite{haldane08a}. Soon after that, many experimental
realizations followed using microwave photons~\cite{wang09a}, photonic
crystals~\cite{malkova09a,chen18a}, coupled waveguides~\cite{rechtsman13b} or
resonators~\cite{hafezi11a,zhao18a,parto18a}, exciton-polaritons~\cite{jean17a} or
metamaterials~\cite{chen14a}, to name a few (see~\cite{ozawa19a} and references 
therein for an authoritative review). Nowadays, topological photonics is
a burgeoning field with many experimental and theoretical developments. Among them, 
one of the current frontiers of the field is the exploration of the interplay
between topological photons and quantum
emitters~\cite{perczel17a,bettles17a,barik18a}.  

\begin{figure}[!tb]
	\centering
	\includegraphics[width=\linewidth]{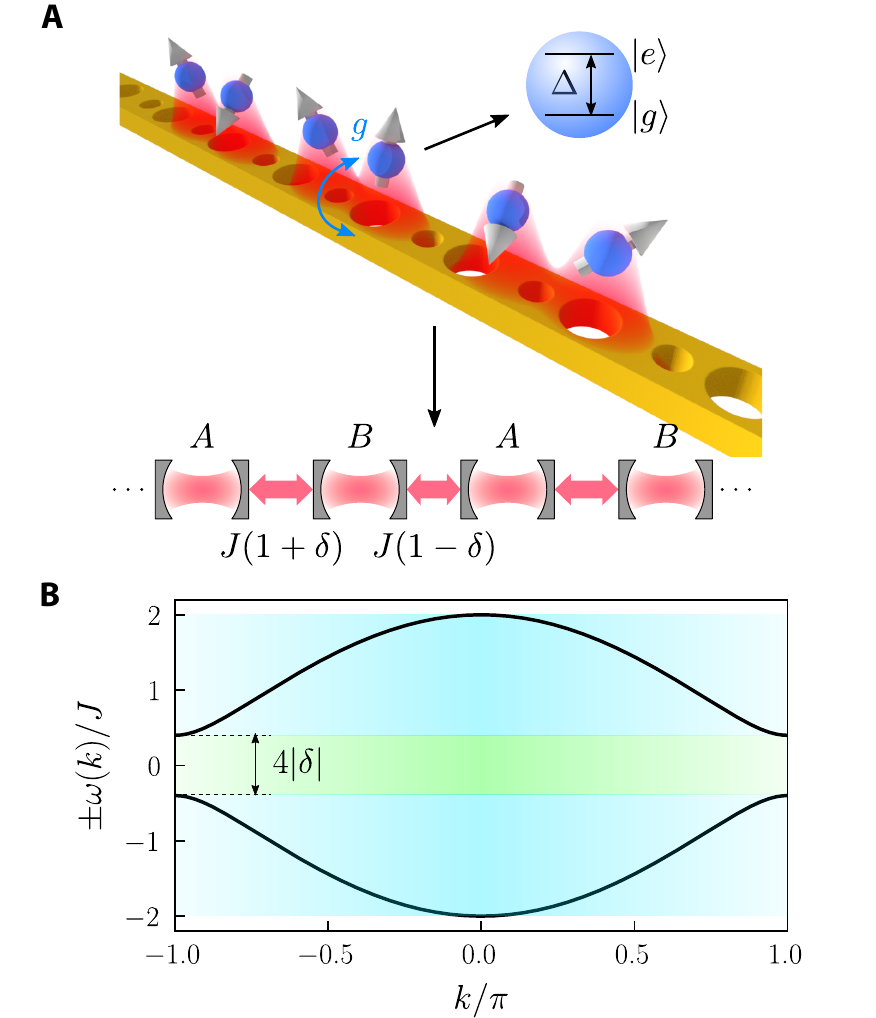}
  \caption{\textbf{System schematic}. (\textbf{A}) Schematic picture of the
  present setup: $N_e$ two-level quantum emitters interact with the photonic
  analogue of the SSH model. This model is characterized by having alternating
  hopping amplitudes $J(1\pm \delta)$, where $J$ defines their strength, while
  $\delta$, the so-called dimerization parameter, controls the asymmetry between
  them. The interaction with photons (in transparent red) induces non-trivial
  dynamics between the emitters.  
  (\textbf{B}) Bath's energy bands for a system with a dimerization parameter
  $|\delta|=0.2$. The main spectral regions of interest for this manuscript are
  the middle band-gap (green) and the two bands (blue).}   
  \label{fig:schematic}
\end{figure}

In this manuscript, we show that topological photonic systems cause a number of
unprecedented phenomena in the field of quantum optics, namely, when they are
coupled to quantum emitters. We analyze the simplest model consisting of
two-level quantum emitters (QEs) interacting with a one-dimensional
topological photonic bath described by the Su-Schrieffer-Heeger (SSH)
model~\cite{asbothbook16a} (see Fig.~\ref{fig:schematic}). When the QE frequency
lies between the two bands (green region in Fig.~\ref{fig:schematic}B) we
predict the emergence of \emph{chiral} photon bound states (BS), that is, BSs
which localize to the left/right side of the QEs depending on the topology of
the bath. 
In the many-body regime (i.e., with many emitters)
those BSs mediate tunable, chiral, long-range interactions, leading to a rich
phase diagram at zero temperature, e.g., with double N\'eel-ordered phases.
Furthermore, when the QEs are resonant with the bands (blue regions in
Fig.~\ref{fig:schematic}B), we also find unusual dissipative dynamics. For
example, for two equal QEs separated a given distance, we show that both the
super/subradiance conditions~\cite{dicke54a} and the scattering properties
depend on the parameter that governs the bath topology even though the energy
dispersion $\omega(k)$ is insensitive to it. This might open avenues to probe 
the topology of these systems in unconventional ways, e.g., through
reflection/transmission experiments. 


\section{Light-matter interactions with one-dimensional topological 
baths~\label{sec:lightmatter}}

The system that we study along this manuscript is sketched in
Fig.~\ref{fig:schematic}A: one or many QEs interact through a common bath which
behaves as the photonic analogue of the SSH model~\cite{asbothbook16a}. This bath model
is described by two interspersed photonic lattices $A$/$B$ of size $N$ with
alternating nearest neighbour hoppings $J(1\pm\delta)$ between their photonic
modes. Assuming periodic boundary conditions and defining
$V^\dagger=(a^\dagger_k,b_k^\dagger)$, the bath Hamiltonian can be written in
momentum space as $H_B=\sum_k V^\dagger \tilde{H}_B(k) V$, with (setting
$\hbar=1$): 
\begin{equation}
  \tilde{H}_B(k) = \left( \begin{array}{cc}
          \omega_a & f(k) \\
            f^*(k) & \omega_a
  \end{array} \right)\,,
  \label{eq:HB}
\end{equation}
where $f(k)=-J\left[(1+\delta)+(1-\delta)e^{-ik}\right]=\omega(k)e^{i\phi(k)}$
(with $\omega(k)>0$) is the coupling in momentum space between the $A$ ($B$)
modes, $a_k = \sum_j a_j e^{-ik j}/\sqrt{N}$ ($b_k = \sum_j b_j e^{-ik
j}/\sqrt{N}$). Here $a_j^\dagger/b_j^\dagger$ ($a_j/b_j$) are the creation
(annihilation) operators of the $A$/$B$ photonic mode at the $j$'th unit cell.
We assume that the $A$/$B$ modes have the same energy, $\omega_a$, that from
now on will be the reference energy of the problem, i.e., $\omega_a\equiv 0$.
This Hamiltonian can be easily diagonalized introducing the eigenoperators,
$u_k/l_k=\left[\pm a_k+e^{i\phi(k)} b_k\right]/\sqrt{2}$, as $H_B=\sum_k
\omega_k(u_k^\dagger u_k-l^\dagger_k l_k)$, leading to two bands with energy:
\begin{equation}
\pm \omega(k)=\pm J\sqrt{2(1+\delta^2)+2(1-\delta^2)\cos(k)}\,.
\end{equation}

Let us now summarize the main bath properties:
\begin{itemize}
	\item The bath possesses sublattice (chiral) symmetry~\cite{asbothbook16a},
    such that all eigenmodes can be grouped in chiral-symmetric pairs with
    opposite energies. Thus, the two bands are symmetric with respect to
    $\omega_a$, spanning from $[-2J,-2|\delta|J]$ (lower band) and
    $[2|\delta|J,2J]$ (upper band). The middle gap is $4|\delta|J$, such that it
    closes when $\delta=0$, recovering the normal 1D tight-binding model. 
	
	\item This bath supports topologically non-trivial phases,  belonging to the
    BDI class in the topological classification of phases~\cite{ryu10a}. More
    concretely, both bands can be characterized by a topological invariant, the
    Zak phase~\cite{asbothbook16a} $\mathcal{Z}$, such that $\mathcal{Z}=0$
    corresponds to a trivial insulator, while $\mathcal{Z}=\pi$ implies a
    non-trivial insulator. For the parametrization we have chosen this occurs for
    $\delta>0$ and $\delta<0$, respectively. Notice that for an infinite system
    (i.e., in the bulk), this definition depends on the choice of the unit cell
    and the role of $\delta$ can be reversed by shifting the unit cell
    by one site. In the bulk the band topology manifests in the fact that
    one can not transform from one phase to the other without closing the gap
    (as long as the symmetry is preserved).  

  \item With finite systems, however, the sign of $\delta$ determines whether
    the chain ends with weak/strong hoppings, which leads to the appearance
    (or not) of topologically robust edge states~\cite{ryu02a}.  
\end{itemize}

Now, let us finally describe the rest of the elements of our setup. For the $N_e$ QEs,
we consider they all have a single optical transition $g$-$e$, with a detuning
$\Delta$ respect to $\omega_a$, and they couple to
the bath locally. Thus, their free and interaction Hamiltonian read: 
\begin{align}
H_S&= \Delta\sum_{m=1}^{N_e}\sigma^m_{ee}\,,\\
H_I&= g\sum_m\left(\sigma^{m}_{eg}c_{x_m}+\mathrm{H.c.}\right) \,, \label{eq:HI}
\end{align}
where $c_{x_m}\in\{a_{x_m},b_{x_m}\}$ depends on the sublattice and the unit
cell $x_m$ at which the $m$'th QEs couples to the bath. We use the notation
$\sigma_{\mu\nu}^m=\ket{\mu}_m\bra{\nu}$, $\mu,\nu\in\{e, g\}$, for the $m$'th
QE operator. We highlight that we use a rotating-wave approximation, such that
only excitation-conserving terms appear in $H_I$.  

\emph{Methods.}
In the next sections, we study the dynamics emerging from the global
QE-bath Hamiltonian $H=H_S+H_B+H_I$ using several complementary approaches.
When one is only interested in the QE dynamics, and the bath can be effectively
traced out, the following Born-Markov master equation~\cite{gardiner_book00a}
describes the evolution of the reduced density matrix $\rho$ of the QEs: 
\begin{multline}
  \dot{\rho} = i[\rho,H_S] + 
  i\sum_{n,m}J_{mn}^{\alpha\beta}\left[\rho,\sigma_{eg}^m \sigma_{ge}^{n}\right] \\
  + \sum_{n,m}\frac{\Gamma^{\alpha\beta}_{mn}}{2}\left[2\sigma_{ge}^n\rho\sigma_{eg}^m 
  - \sigma_{eg}^m\sigma_{ge}^n\rho -\rho\sigma_{eg}^m\sigma_{ge}^n\right]\,.
  \label{eq:BMMaster}
\end{multline}

The functions $J_{mn}^{\alpha\beta},\Gamma^{\alpha\beta}_{mn}$, which ultimately
control the QE coherent and dissipative dynamics, respectively, are the real and
imaginary part of the collective self-energy
$\Sigma_{mn}^{\alpha\beta}(\Delta+i0^+)=J_{mn}^{\alpha\beta}-i\frac{\Gamma^{\alpha\beta}_{mn}}{2}$.
This collective self-energy depends on the sublattices $\alpha, \beta
\in\{A,B\}$ to which the $m$'th and $n$'th QE couple respectively, as well as on 
their relative position $x_{mn}=x_n-x_m$. Remarkably, for our model they can be
calculated analytically in the thermodynamic limit ($N\rightarrow\infty$) yielding:
\begin{gather}
\Sigma^{AA/BB}_{mn}(z)=-\frac{g^2 z\left[y^{|x_{mn}|}_+\Theta_{+}(y_+)
	-y^{|x_{mn}|}_-\Theta_{-}(y_{+})\right]}
{\sqrt{z^4-4J^2(1+\delta^2)z^2+16J^4\delta^2}} \,, \label{eq:SigmaAA}\\
\Sigma^{AB}_{mn}(z)=\frac{g^2J\left[F_{x_{mn}}(y_+)\Theta_{+}(y_+)-
	F_{x_{mn}}(y_-)\Theta_{-}(y_+)\right]}
{\sqrt{z^4-4J^2(1+\delta^2)z^2+16J^4\delta^2}}\label{eq:SigmaAB} \,,
\end{gather}
where $F_n(z)=(1+\delta)z^{|n|}+(1-\delta)z^{|n+1|}$,
$\Theta_{\pm}(z)={\Theta(\pm 1 \mp|z|)}$, $\Theta(z)$ is Heaviside's step
function, and 
\begin{multline}
  y_\pm=\frac{1}{2J^2(1-\delta^2)}\Big[z^2-2J^2(1+\delta^2) \\ 
  \pm\sqrt{z^4-4J^2(1+\delta^2)z^2+16J^4\delta^2}\Big] \,.
  \label{eq:ypm}
\end{multline} 

However, since we have a highly structured bath this perturbative description
will not be valid in certain regimes, e.g., close to band-edges, and we will use
resolvent operator techniques~\cite{cohenbook92a} or fully numerical
approaches to solve the problem exactly for infinite/finite bath sizes,
respectively. Since those methods were explained in detail in other works, here we focus on the results and leave the details for the Supp.\ Material.  

\section{Band-gap regime~\label{sec:coherent}}

\begin{figure}
  \centering\includegraphics[width=\linewidth]{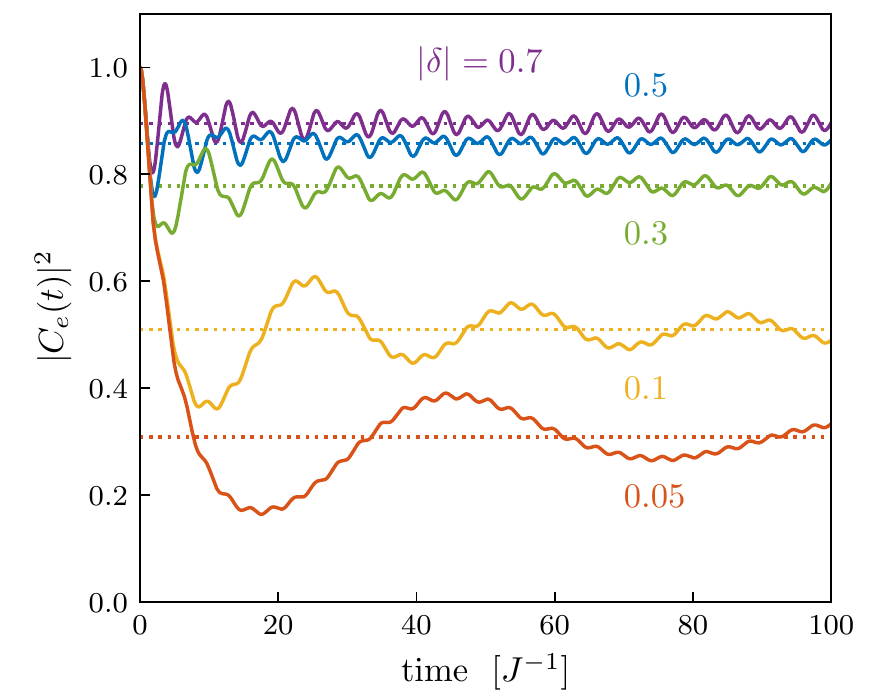}
  \caption{\textbf{Single QE dynamics}. 
    Probability to find the emitter in the excited state, $|C_e(t)|^2$, for
    different values of $|\delta|$. The other parameters are $\Delta=0$ (middle
    of the band gap) and $g=0.4J$. As the band gap closes, i.e.,
    $\delta\rightarrow 0$, the decay becomes stronger. Dashed lines mark the
    value of ${|C_e(t\to\infty)|^2}=\big(1+\frac{g^2}{4J^2|\delta|}\big)^{-2}$.}
	\label{fig:singleQEdynamics}
\end{figure}

In this Section we assume that the QEs are in the band-gap regime, that is,
their transition frequency lies outside of the two bands of the photonic bath.
From here on, we only discuss results in the thermodynamic limit (when
$N\rightarrow\infty$) such that the edge states~\cite{ryu02a} play no role in
the QE dynamics. We refer the interested reader to
Refs.~\cite{ciccarello11a} and Supp.\ Material to see some of the
consequences the edge states have on the QE dynamics.  

\subsection{Single QE: dynamics}

Let us start considering the dynamics of a single excited QE, i.e.,
$\ket{\psi(0)}=\ket{e}\ket{\mathrm{vac}}$, where $\ket{\mathrm{vac}}$ denotes
the vacuum state of the lattice of bosonic modes. Since $H$ conserves the number
of excitations, the global wavefunction at any time reads: 
\begin{equation}
  \ket{\psi(t)}=\left[C_e(t)\sigma_{eg} 
  +\sum_{j=1}^N\sum_{\alpha=a,b}C_{j,\alpha}(t)\alpha^\dagger_j\right]
  \ket{g}\ket{\mathrm{vac}}\,. 
  \label{eq:wavefunction}
\end{equation}

In both perturbative and exact treatments, the dynamics of $C_e(t)$ can be shown
(see Refs.~\cite{cohenbook92a} and Supp.\ Material) to depend only on
the single QE self-energy:
\begin{equation}
  \Sigma_e(z)= \frac{g^2z\,\mathrm{sign}(|y_+|-1)}{\sqrt{z^4-4J^2(1+\delta^2)z^2+16J^4\delta^2}} \,, 
  \label{eq:singleQEselfenergy}
\end{equation}
obtained from Eq.~\ref{eq:SigmaAA} defining: $\Sigma_{e}(z)\equiv
\Sigma_{nn}^{AA}(z)$. From here, we can already extract several conclusions: i)
$\Sigma_e(z)$ is independent of the sign of $\delta$, which means that the
spontaneous emission dynamics is insensitive to the topology of the bands. ii)
Perturbative approaches, like the Born-Markov approximation of
Eq.~\ref{eq:BMMaster}, predict an exponential decay of excitations at a rate
$\Gamma_e(\Delta)=-2\mathrm{Im}\Sigma_e(\Delta+i0^+)$, which is strictly zero in
the band-gap regime. Thus, one expects that the excitation remains localized in
the QE at any time. However, in Fig.~\ref{fig:singleQEdynamics} we compute the
exact dynamics $C_e(t)$ for several $\delta$'s and observe that this
perturbative limit is only recovered in the limit of $|\delta|\to 1$. On
the contrary, when $|\delta|\ll 1$ and $\delta\neq 0$ the dynamics displays
fractional decay and oscillations. As it happens with other
baths~\cite{john94a},
the origin of this dynamics stems from the emergence of photon bound
states (BSs) which localize around the
QEs~\cite{bykov75a,john90a,kurizki90a}. However, the
BSs appearing in the present topological waveguide bath have some distinctive features
with no analogue in other systems, and therefore deserve special attention. 

\begin{figure*}
  \centering\includegraphics[width=\linewidth]{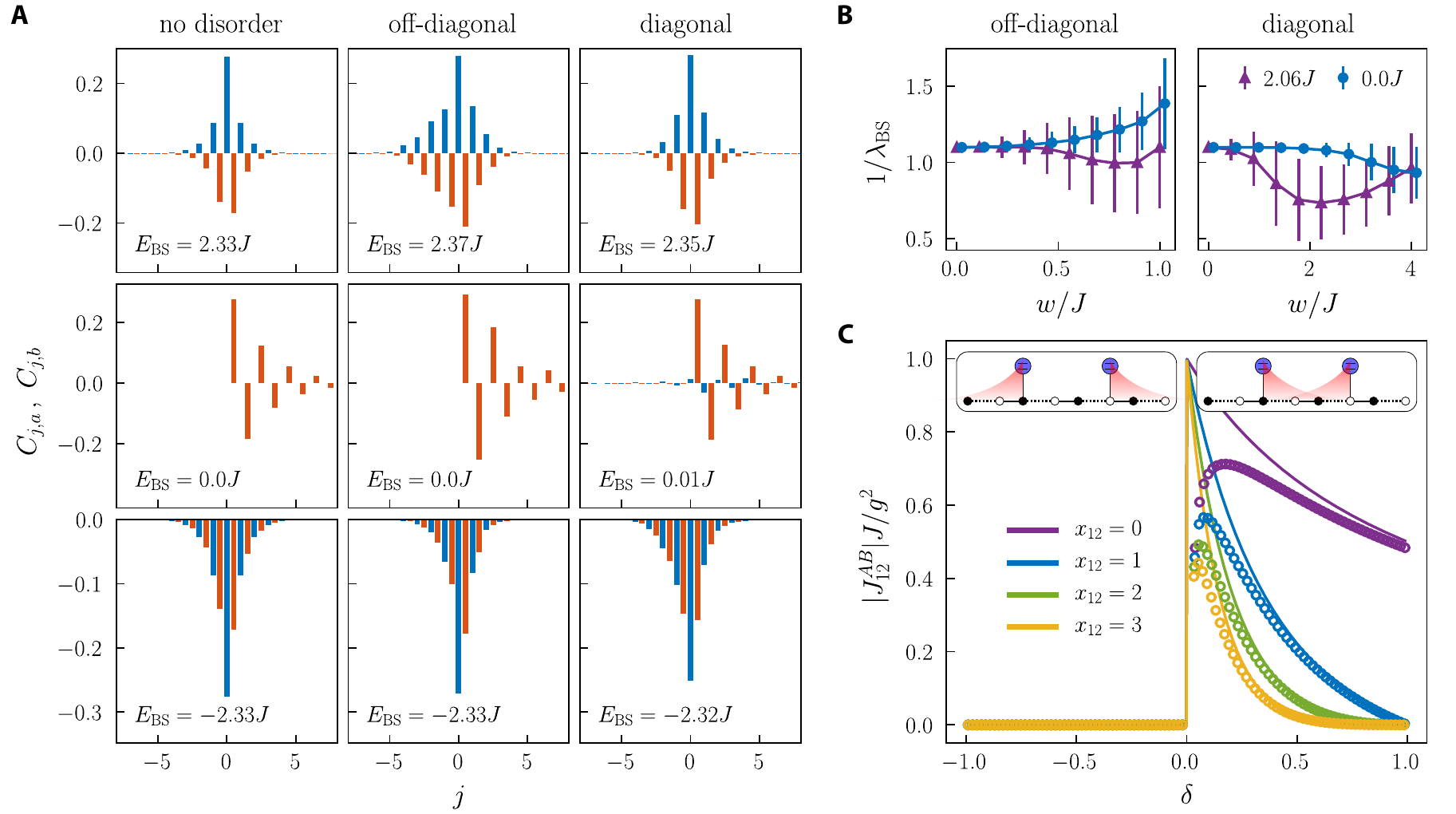}
  \caption{\textbf{Bound state properties.} 
  (\textbf{A}) BS wavefunction for a QE placed at $j=0$ that couples to the $A$
  sublattice; $\delta=0.2$ and $g=0.4J$. Probability amplitudes $C_{j,a}$ are
  shown in blue, while the $C_{j,b}$ are shown in red. The QE frequency is set
  to $\Delta=2.2J$ (top row), $\Delta=0$ (middle row) and $\Delta=-2.2J$ (bottom
  row). The first column corresponds to the model without disorder, the second
  to the model with disorder in the couplings between cavities and the third to
  the model with disorder in the cavities resonant frequencies. In both cases
  with disorder its strength is set to $w=0.5J$. For each case, the value of the
  bound state's energy is shown at the bottom of the plots.  
  (\textbf{B}) Inverse bound state localization length for the two different
  models of disorder as a function of the disorder strength. Parameters:
  $g=0.4J$, $\delta=0.5$. The dots correspond to the average value computed with a
  total of $10^4$ instances of disorder, and the error bars mark the value of
  one standard deviation above and below the average value (the blue curves are
  slightly displaced to the right for better visibility). Two cases are shown
  which correspond to $\Delta\simeq 2.06J$ (triangles, outer band gaps), and 
  $\Delta=0$ (circles, inner band gap).   
  (\textbf{C}) Absolute value of the dipolar coupling for $\Delta=0$ and
  $g=0.4J$; Markov (solid line), exact (dots). The insets show the shape of the
  bound states in the topological and the trivial phases. The situation for the
  $BA$ configuration is the same, reversing the role of $\delta$.} 
  \label{fig:boundstates}
\end{figure*}

\subsection{Single QE: Bound states}

The energy and wavefunction of the BSs in the single-excitation subspace can be
obtained by solving the secular equation
$H\ket{\Psi_\mathrm{BS}}=E_\mathrm{BS}\ket{\Psi_\mathrm{BS}}$, with
$E_\mathrm{BS}$ lying out of the bands, and $\ket{\Psi_\mathrm{BS}}$ in the form
of Eq.~\eqref{eq:wavefunction}, but with time-independent coefficients. Without
loss of generality, we assume that the QE couples to sublattice $A$ at the $j=0$
cell. After some algebra, one can find that the energy of the BS is given by the
pole equation: $E_{\mathrm{BS}}=\Delta+\Sigma_e(E_\mathrm{BS})$. Irrespective of
$\Delta$ or $g$, there exist always three BS solutions of the pole equation (one
for each band-gap region). This is because the self-energy diverges in all
band-edges, which guarantees finding a BS in each of the
band-gaps~\cite{calajo16a,shi16a}. 
The main difference with respect to other
BSs~\cite{bykov75a,john90a,kurizki90a,calajo16a,shi16a}
appears in the wavefunction amplitudes, which read 
\begin{gather}
  C_{j,a}=\frac{gE_\mathrm{BS}C_e}{2\pi}\int_{-\pi}^\pi dk
  \frac{e^{ikj}}{E^2_\mathrm{BS}-\omega^2(k)} \,, \label{eq:wavea} \\
  C_{j,b}=\frac{gC_e}{2\pi}\int_{-\pi}^\pi dk
  \frac{\omega(k)e^{i[kj-\phi(k)]}}{E^2_\mathrm{BS}-\omega^2(k)} \,,
  \label{eq:waveb} 
\end{gather}
where $C_e$ is a constant obtained from the normalization condition that is
directly related with the long-time population of the excited state in
spontaneous emission. For example, in Fig.~\ref{fig:singleQEdynamics} where
$\Delta=0$, it can be shown to be 
$|C_e(t\to \infty)|^2=|C_e|^4=\big(1+\frac{g^2}{4J^2|\delta|}\big)^{-2}$.

From Eqs.~\ref{eq:wavea}-\ref{eq:waveb}, we can extract several properties
of the spatial wavefunction distribution. On the one hand, above or
below the bands (outer band gaps) the largest contribution to the integrals is
that of $k=0$, thus all the $C_{j,\alpha}$ have the same sign (see left column of
Fig.~\ref{fig:boundstates}A top and bottom row). In the lower (upper) band-gap,
$C_{j,\alpha}$ of the different sublattices has the same (opposite) sign. On the
other hand, in the inner band gap, the main contribution to the integrals is
that of $k=\pi$. This gives an extra factor $(-1)^j$ to the coefficients $C_{j,\alpha}$
(see Fig.~\ref{fig:boundstates}A middle row). Furthermore, the probability
amplitudes of the sublattice which the QE couples to are symmetric with respect
to the position of the QE, whereas they are asymmetric in the other sublattice,
that is, the BSs are chiral. Changing $\delta$ from positive to negative results
in a spatial inversion of the BS wavefunction. The asymmetry of the BS
wavefunction is more extreme in the middle of the band-gap ($\Delta=0$). For example, 
if $\delta>0$, the BS wavefunction with $E_\mathrm{BS}=0$ is given by
$C_{j,a}=0$ and 
\begin{equation}
  C_{j,b}=\begin{cases}
  \frac{gC_e(-1)^j}{J(1+\delta)}\left(\frac{1-\delta}{1+\delta}\right)^j\,, 
    & j\geq 0 \\
  0\,, & j<0 
  \end{cases} \ ,
\end{equation}
whereas for $\delta<0$ the wavefunction decays for $j<0$ while being strictly
zero for $j\geq 0$. At this point, the BS decay length diverges as $\lambda_\mathrm{BS}\sim
1/(2|\delta|)$ when the gap closes. Away from this point, the BS decay length
shows the usual behavior for 1D baths $\lambda_\mathrm{BS}\sim 1/\sqrt{|\Delta_\mathrm{edge}|}$,
with $\Delta_\mathrm{edge}$ being the smallest detuning between the QE and the band-edge
frequencies.

The physical intuition of the appearance of such chiral BS at
${E_\mathrm{BS}=0}$ is that the QE with $\Delta=0$ acts as an \textit{effective
edge} in the middle of the chain, or equivalently, as a boundary between two
semi-infinite chains with different topology. This picture provides us with an insight
useful to understand other results of the manuscript: despite considering the
case of an infinite bath, the local QE-bath coupling inherits information about the
underlying bath topology. In fact, one can show that this chiral BS has the same
properties as the edge-state which appears in a semi-infinite SSH chain in the
topologically non-trivial phase, for example, inheriting its robustness to
disorder. To illustrate it, we study the effect of two types of disorder: one
that appears in the cavities bare frequencies (diagonal), and another one that
appears in the tunneling amplitudes between them (off-diagonal). The former
corresponds to the addition of random diagonal terms to the bath's Hamiltonian
$H_B\to H_B + \sum_j\left(\epsilon_{a,j}a^\dagger_ja_j +
\epsilon_{b,j}b^\dagger_jb_j\right)$ and breaks the chiral symmetry of the
original model, while the latter corresponds to the addition of off-diagonal
random terms $H_B\to H_B + \sum_j\left(\epsilon_{1,j}b^\dagger_ja_j +
\epsilon_{2,j}a^\dagger_{j+1}b_j + \mathrm{H.c.}\right)$ and preserves it. We
take the $\epsilon_{\nu,j}$, $\nu=a,b,1,2$, from a uniform distribution within
the range $[-w/2,w/2]$ for each $j$'th unit cell. To prevent changing
the sign of the coupling amplitudes between the cavities, $w$ is restricted to
$w/2 < (1-|\delta|)$ in the case of off-diagonal disorder.

In the middle~(right) column of Fig.~\ref{fig:boundstates}A we plot the shape of
the three BS appearing in our problem for a situation with off-diagonal
(diagonal) disorder with $w=0.5J$. There, we observe that while the upper and lower
BS get modified for both types of disorder, the chiral BS has the same
protection against off-diagonal disorder as a regular SSH edge-state: its
energy is pinned at $E_\mathrm{BS}=0$ as well as keeping its shape with no
amplitude in the sublattice to which the QE couples to. On the contrary, for
diagonal disorder the middle BS is not protected any more and may have weight in
both sublattices.  

Finally, to make more explicit the different behaviour with disorder of the
middle BS compared to the other ones, we compute their localization length
$\lambda_{\mathrm{BS}}$ as a function of the disorder strength $w$ averaging for
many realizations. In Fig.~\ref{fig:boundstates}B we plot both the average value
(markers) of $\lambda_\mathrm{BS}^{-1}$ and its standard deviation (bars) for
the cases of the middle (blue circles) and upper (purple triangles) BSs.
Generally, one expects that for weak disorder, states outside the band regions
tend to delocalize, while for strong disorder all eigenstates become localized
(see, for example, Ref.~\cite{Bea2018}). In fact, this is the behaviour we
observe for the upper BS for both types of disorder. However, the numerical
results suggest that for off-diagonal disorder the chiral BS never delocalizes
(on average). Furthermore, the chiral BS localization length is less sensitive
to the disorder strength $w$ manifested in both the large initial plateau region
as well as the smaller standard deviations compared to the upper BS results.  

Summing up, a QE coupled locally to an SSH bath: i) localizes a photon only at
one side of the emitter depending on the sign of $\delta$, ii) with no amplitude
in the sublattice where the QE couples to, iii) with the same properties as the
topological edge states, e.g., robustness to disorder. As we discuss in more
detail in the Supp.\ Material, the SSH bath is the simplest one-dimensional bath
that provides all these features simultaneously.

\subsection{Two QEs}

Let us now focus on the consequences of such exotic BS when two QEs are coupled
to the bath. For concreteness, we focus on a parameter regime where the
Born-Markov approximation is justified, although we have performed an exact
analysis in the Supp.\ Material. From Eq.~\ref{eq:BMMaster}, it is easy to see
that in the band-gap regime, the interaction with the bath leads to an effective
unitary dynamics governed by the following Hamiltonian:
\begin{equation}
  H_\mathrm{dd}=J^{\alpha\beta}_{12}\left(\sigma^{1}_{eg}\sigma^{2}_{ge} 
  + \mathrm{H.c.}\right) \,.  
  \label{eq:spinH}
\end{equation}
That is, the bath mediates dipole-dipole interactions between the QEs. One way
to understand the origin of these interactions is that the emitters exchange
virtual photons through the bath, which in this case are localized around the emitter. In fact, these virtual photons are nothing but the photon BS that we studied in the previous Section. Thus, these interactions 
$J^{\alpha\beta}_{mn}$ inherit many properties of the BSs. For example,
the interactions are exponentially localized in space, with a localization
length that can be tuned and made large by setting $\Delta$ close to the band-edges,
or fixing $\Delta=0$ and letting the middle band-gap close ($\delta\rightarrow
0$). Moreover, one can also change qualitatively the interactions by moving
$\Delta$ to different band-gaps: for $|\Delta|>2J$ all the
$J^{\alpha\beta}_{mn}$ have the same sign, while for $|\Delta|<2|\delta|J$ they
alternate sign as $x_{mn}$ increases. Also, changing $\Delta$ from
positive to negative changes the sign of $J^{AA/BB}_{mn}$, but leaves unaltered
$J^{AB/BA}_{mn}$. Furthermore, while $J^{AA/BB}_{mn}$ are insensitive to the
bath's topology, the $J^{AB/BA}_{mn}$ mimic the dimerization of the underlying
bath, but allowing for longer range couplings.  The most striking regime is again 
reached for $\Delta=0$. In that case
$J^{AA/BB}_{mn}$ identically vanish, and thus the QEs only interact if they
are coupled to different sublattices. Furthermore, in such a situation the
interactions have a strong directional character, i.e., the QEs only
interact if they are in some particular order. Assuming that the first QE
at $x_1$ couples to sublattice $A$, and the second one at $x_2$ couples to
$B$, we have 
\begin{equation}
  J^{AB}_{12}=\begin{cases}
  \mathrm{sign}(\delta)\frac{g^2(-1)^{x_{12}}}{J(1+\delta)}\left(\frac{1-\delta}{1+\delta}\right)^{x_{12}}
    & \text{if } \delta \cdot x_{12} > 0 \,. \\
    0 & \text{if } \delta \cdot x_{12} < 0 \,. \\
    \Theta(\delta)\frac{g^2}{J(1+\delta)} & \text{if } x_{12} = 0 \,. 
  \end{cases}
  \label{eq:JAB_middle}
\end{equation}
In Fig.~\ref{fig:boundstates}C we plot the absolute value of the coupling for
this case computed exactly, and compare it with the Markovian formula. Apart
from small deviations at short distances, it is important to highlight that the
directional character agrees perfectly in both cases. 

\subsection{Many QEs: Spin models with topological long-range interactions}

\begin{figure*}
  \centering\includegraphics[width=\linewidth]{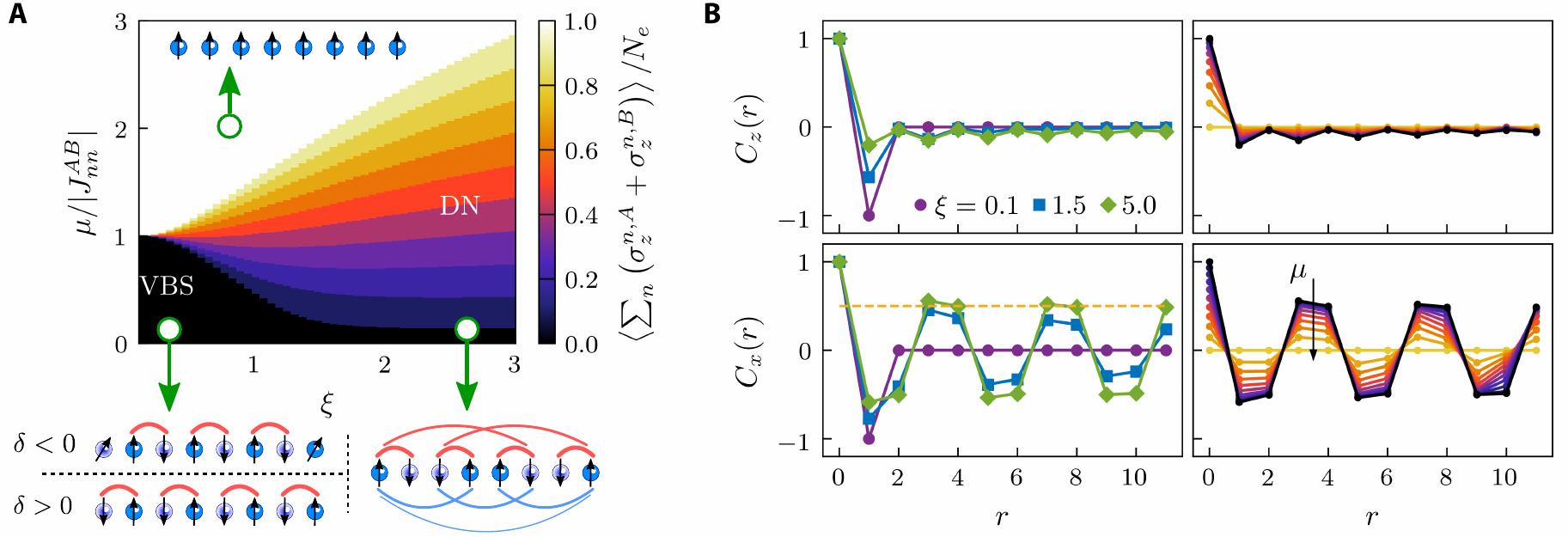}
  \caption{\textbf{Spin models: phase diagram and correlations}. (\textbf{A})
  Ground state average polarization obtained by exact diagonalization 
  for a chain with $N_e=20$ emitters with frequency
  tuned to $\Delta=0$ as a function of the chemical
  potential $\mu$ and the decay length of the interactions $\xi$. The different
  phases discussed in the text, a Valence-Bond Solid (VBS) and a Double N\'eel
  ordered phase (DN) are shown schematically below, on the left
  and right respectively. Interactions of different sign are marked with links
  of different color.
  For the VBS we show two possible configurations
  corresponding to $\delta<0$ (top) and $\delta>0$ (bottom). In the
  topologically non-trivial phase ($\delta<0$) two spins are left uncoupled with
  the rest of the chain. 
  (\textbf{B}) Correlations ${C_\nu(r) = \mean{\sigma^{9}_\nu
  \sigma^{9+r}_\nu} - \mean{\sigma^{9}_\nu}\mean{\sigma^{9+r}_\nu}}$,
  $\nu=x,y,z$ [$C_x(r)=C_y(r)$] for the same system as in (A) for different
  interaction lengths, fixing $\mu=0$ (left column). Correlations for different 
  chemical potentials fixing $\xi=5$, darker colors correspond to lower chemical 
  potentials (right column). Note we have defined a single index $r$ that
  combines the unit cell position and the sublattice index. The yellow dashed 
  line marks the value of $1/2$ expected when the interactions are of infinite
  range.} 
	\label{fig:spinorders}
\end{figure*}

One of the main interests of having a platform with BS-mediated interactions is
to investigate spin models with long-range
interactions~\cite{douglas15a,gonzaleztudela15c}. The study of these
models has become an attractive avenue in quantum simulation because long-range
interactions are the source of non-trivial many-body
phases~\cite{hauke10c} and
dynamics~\cite{richerme14a}, and are also very hard to treat
classically.  

Let us now investigate how the shape of the QE interactions inherited from the
topological bath translate into different many-body phases at zero temperature
as compared to those produced by long-range interactions appearing in other setups such as
trapped ions~\cite{hauke10c,richerme14a}, or standard waveguide setups. For that, we
consider having $N_e$ emitters equally spaced and alternatively coupled to the
$A$/$B$ lattice sites. After eliminating the bath, and adding a collective field
with amplitude $\mu$ to control the number of spin excitations, the dynamics of the
emitters (spins) is effectively given by: 
\begin{multline}
H_\mathrm{spin} = \sum_{m,n} \left[J^{AB}_{mn}\left(
\sigma_{eg}^{m,A}\sigma_{ge}^{n,B}+\mathrm{H.c.}\right)\right. \\
\left. - \frac{\mu}{2} \left(\sigma_{z}^{m,A} + \sigma_{z}^{n,B}\right)
\right] \,,
\label{eq:Hspinmiddle}
\end{multline}
denoting by $\sigma_{\nu}^{n,\alpha}$, $\nu=x, y, z$, the corresponding Pauli
matrix acting on the $\alpha\in\{A, B\}$ site in the $n$'th unit cell. The
$J^{\alpha\beta}_{mn}$ are the spin-spin interactions derived in the previous
subsection, whose localization length, denoted by $\xi$, and functional form can
be tuned through system parameters such as $\Delta$.

For example, when the lower (upper) BS mediates the interaction, the $J^{\alpha\beta}_{mn}$
has negative (alternating) sign for all sites, similar to the ones appearing in
standard waveguide setups. When the range of the
interactions is short (nearest neighbor), the physics is well described by the
ferromagnetic XY model with a transverse field~\cite{Katsura1962}, which goes
from a fully polarized phase when $|\mu|$ dominates to a superfluid one in which
spins start flipping as $|\mu|$ decreases. In the case where the interactions
are long-ranged the physics is similar to that explained in Ref.~\cite{hauke10c}
for power-law interactions ($\propto 1/r^3$). The longer range of the
interactions tends to break the symmetry between the ferro/antiferromagnetic
situations and leads to frustrated many-body phases. Since similar
interactions also appear in other scenarios (standard waveguides or trapped
ions), we now focus on the more different situation where the middle BS at
$\Delta=0$ mediates the interactions, such that the coefficients $J^{AB}_{mn}$ have
the form of Eq.~\eqref{eq:JAB_middle}.  

In that case, the Hamiltonian $H_\mathrm{spin}$ of Eq.~\ref{eq:Hspinmiddle} is
very unusual: i) spins only interact if they are in different sublattices, i.e.,
the system is bipartite ii)
the interaction is chiral in the sense that they interact only in case they
are properly sorted: the one in lattice $A$ to the left/right of that in lattice
$B$, depending on the sign of $\delta$. Note that $\delta$ also controls the
interaction length $\xi$.  In particular, for $|\delta|=1$ the interaction only
occurs between nearest neighbors, whereas for $\delta\to 0$, the interactions
become of infinite range. These interactions translate into a rich phase diagram
as a function of $\xi$ and $\mu$, which we plot in Fig.~\ref{fig:spinorders}A
for a small chain with $N_e=20$ emitters (obtained with exact diagonalization).
Let us guide the reader into the different parts: 
\begin{itemize}
	\item The region with maximum average magnetization (in white) corresponds to
    the places where $\mu$ dominates such that all spins are alligned upwards.

	\item Now, if we decrease $\mu$ from this fully polarized phase in a region
    where the localization length is short, i.e., $\xi\approx 0.1$, we observe a
    transition into a state with zero average magnetization. This behaviour can
    be understood because in that short-range limit $J^{AB}_{mn}$ only couples
    nearest neighbor $AB$ sites, but not $BA$ sites as shown in
    the scheme of the lower part of the diagram for $\delta>0$ (the opposite is
    true for $\delta<0$). Thus, the ground state is a
    product of nearest neighbor singlets (for $J>0$) or triplets (for $J<0$).
    This state is usually referred to as Valence-Bond Solid in the condensed
    matter literature~\cite{auerbach12a}. 
    Note, the difference between $\delta \gtrless 0$ is the presence (or not) of
    uncoupled spins at the edges.  

	\item However, when the bath allows for longer range interactions
    ($\xi>1$), the transition from the fully polarized phase to the
    phase of zero magnetization does not occur abruptly but passing through all
    possible intermediate values of the magnetization. Besides, we also plot in
    Fig.~\ref{fig:spinorders}B the spin-spin correlations along the $x$ and $z$
    directions (note the symmetry in the $xy$ plane) for the case of $\mu=0$ to
    evidence that a qualitatively different order appears as $\xi$ increases. In
    particular, we show that the spins align along the $x$ direction with a double
    periodicity, which we can pictorially represent by
    $\ket{\up\up\down\down\up\up\dots}_x$, and that we label as double N\'eel
    order states. Such orders have been predicted as a consequence of
    frustration in classical and quantum spin chains with competing nearest and
    next-nearest neighbour interactions~\cite{morita72a,
    Sen1989,Sen1992}, introduced to describe complex solid state
    systems such as multiferroic materials~\cite{Qi2016}. In our case, this
    order emerges in a system which has long-range interactions but no
    frustration as the system is always bipartite regardless the interaction
    length.    

 \end{itemize}
  
To gain analytical intuition of this regime, we take the limit $\xi\rightarrow
\infty$, where the Hamiltonian~\eqref{eq:Hspinmiddle} reduces to 
\begin{equation}
H'_\mathrm{spin}=UH_\mathrm{spin}U^\dagger
\simeq J(S^+_AS^-_B + \mathrm{H.c.}) \,,
  \label{eq:Hinfinite}
\end{equation}
where $S^+_{A/B}=\sum_n \sigma^{n,A/B}_{eg}$, and we have performed a unitary
transformation $U=\prod_{n\in\mathbb{Z}_\mathrm{odd}}
\sigma^{n,A}_z\sigma^{n,B}_z$, to cancel the alternating signs of
$J^{AB}_{mn}$. Equality in Eq.~\eqref{eq:Hinfinite} occurs for a system with periodic boundary
conditions, while for finite systems with open boundary conditions some
corrections have to be taken into account due to the fact that not all spins in
one sublattice couple to all spins in the other but only to those to their
right/left depending on the sign of $\delta$. 
The ground state is symmetric under (independent) permutations in $A$ and $B$. In
the thermodynamic limit we can apply mean field, which predicts symmetry
breaking in the spin $xy$ plane. For instance, if $J<0$ and the symmetry is
broken along the spin direction $x$, the spins will align so that $\mean{(S^x_A)^2}=\mean{(S^x_B)^2}=\mean{S^x_AS^x_B}=(N_e/2)^2$, and 
$\mean{S^x_A}^2 = \mean{S^x_B}^2 = (N_e/2)^2$\,.

Since $N_e$ is finite in our case, the symmetry is not broken, but it is
still reflected in the correlations, so that 
\begin{equation}
\mean{\sigma^{m,A}_\nu \sigma^{n,A}_\nu} \simeq \mean{\sigma^{m,A}_\nu
	\sigma^{n,B}_\nu} \simeq 1/2 \,,
\end{equation}
with $\nu=x,y$. In the original picture with respect to $U$, we obtain the
double N\'eel order observed in Fig.~\ref{fig:spinorders}B. As can be
understood, the alternating nature of the interactions is crucial for obtaining
this type of ordering. Finally, let us
mention that the topology of the bath translates into the topology of the spin
chain in a straightforward manner: regardless the range of the effective
interactions, the ending spins of the chain will be uncoupled to the rest of
spins if the bath is topologically non-trivial. 

This discussion shows the potential of the present setup to act as a quantum
simulator of exotic many-body phases not possible to simulate with other known
setups. The full characterization of such spin models with topological
long-range interactions is interesting on its own 
and we will present it elsewhere.   

\section{Band regime~\label{sec:dissipative}}

Here, we study the situation when QEs are resonant with one of the bands. For
concreteness, we only present two results where the unconventional nature of the
bath plays a prominent role, namely, the emergence of unexpected
super/subradiant states, and their consequences when a single-photon scatters
into one or two QEs.  

\subsection{Dissipative dynamics: super/subradiance~\label{subsec:subrad}}

The band regime is generally characterized by inducing non-unitary dynamics in
the QEs.  However, when many QEs couple to the bath there are situations in
which the interference between their emission may enhance or suppress (even
completely) the decay of certain states. This phenomenon is known as
super/subradiance~\cite{dicke54a}, respectively, and it can be used, e.g., for
efficient photon storage~\cite{asenjogarcia17a} or multiphoton
generation~\cite{gonzaleztudela17c}. Let us illustrate this effect with two QEs:
In that case, the decay rate of a symmetric/antisymmetric combination of
excitations is $\Gamma_e\pm\Gamma_{12}$. When $\Gamma_{12}=\pm\Gamma_e$, these
states decay at a rate that is either twice the individual one or zero. In this
latter case they are called perfect subradiant or dark states.  

In standard one-dimensional baths~$\Gamma_{12}(\Delta)=\Gamma_e(\Delta)
\cos\big(k(\Delta) |x_{mn}|\big)$, so the dark states are such that the wavelength of
the photons involved, $k(\Delta)$, allows for the formation of a standing wave
between the QEs when both try to decay, i.e., when $k(\Delta) |x_{mn}|=n\pi$,
with $n\in \mathbb{Z}$. Thus, the emergence of perfect super/subradiant states solely
depends on the QE frequency $\Delta$, bath energy dispersion $\omega(k)$, and
their relative position $x_{mn}$, which is the common intuition for this
phenomenon.  

This common wisdom gets modified in the bath considered along this manuscript,
where we find situations in which, for the same values of $x_{mn}$, $\omega(k)$ and
$\Delta$, the induced dynamics is very different depending on the sign of
$\delta$. In particular, when two QEs couple to the $A$/$B$ sublattice
respectively, the collective decay reads: 
\begin{equation}
  \Gamma^{AB}_{12}(\Delta)=
  \Gamma_e\mathrm{sign}(\Delta)\cos\big(k(\Delta)x_{12}-\phi(\Delta)\big)\,, \label{eq:colltop}
\end{equation}
which depends both on the photon wavelength mediating the interaction
$k(\Delta)=\arccos\left[\frac{\Delta^2-2J^2(1+\delta^2)}{2J^2(1-\delta^2)}\right]$,
an even function of $\delta$, and on the phase
$\phi(\Delta)\equiv\phi(k(\Delta))$, sensitive to the sign of $\delta$. This
$\phi$-dependence enters through the system-bath coupling when rewriting
$H_\mathrm{I}$ in Eq.~\ref{eq:HI} in terms of the eigenoperators $u_k,l_k$. The
intuition behind it is that even though the sign of $\delta$ does not play a
role in the bath properties of an infinite system, when the QEs couple to it,
the bath embedded between them is different for $\delta\gtrless 0$, making the
two situations inequivalent.  

Using Eq.~\ref{eq:colltop}, we find that to obtain a perfect a super/subradiant
state it must be satisfied: $k(\Delta_s)x_{12}-\phi(\Delta_s)=n\pi$,
$n\in\mathbb{N}$. They come in pairs: If $\Delta_s$ is a superradiant
(subradiant) state in the upper band, $-\Delta_s$ is a subradiant (superradiant)
state in the lower band. In particular, it can be shown that when $\delta<0$, the 
super/subradiant equation has solutions for $n=0,\dots,x_{12}$, while if
$\delta>0$, the equation has solutions for $n=0,\dots,x_{12}+1$. Besides, the 
detunings, $\Delta_s$ at which the
subradiant states appear also satisfy that $J_{12}^{AB}(\Delta_s)\equiv 0$,
which guarantees that these subradiant states survive even in the non-Markovian
regime (with a correction due to retardation which is small as long as
$x_{12}\Gamma_e(\Delta)/(2|v_g(\Delta)|)\ll 1$). Apart
from inducing different decay dynamics, these different conditions for
super/subradiance at fixed $\Delta$ also translate in different
reflection/transmission coefficients when probing the system through photon
scattering, as we show next.  \\

\subsection{Single-photon scattering~\label{subsec:scat}}
\begin{figure}
	\centering\includegraphics[width=0.99\linewidth]{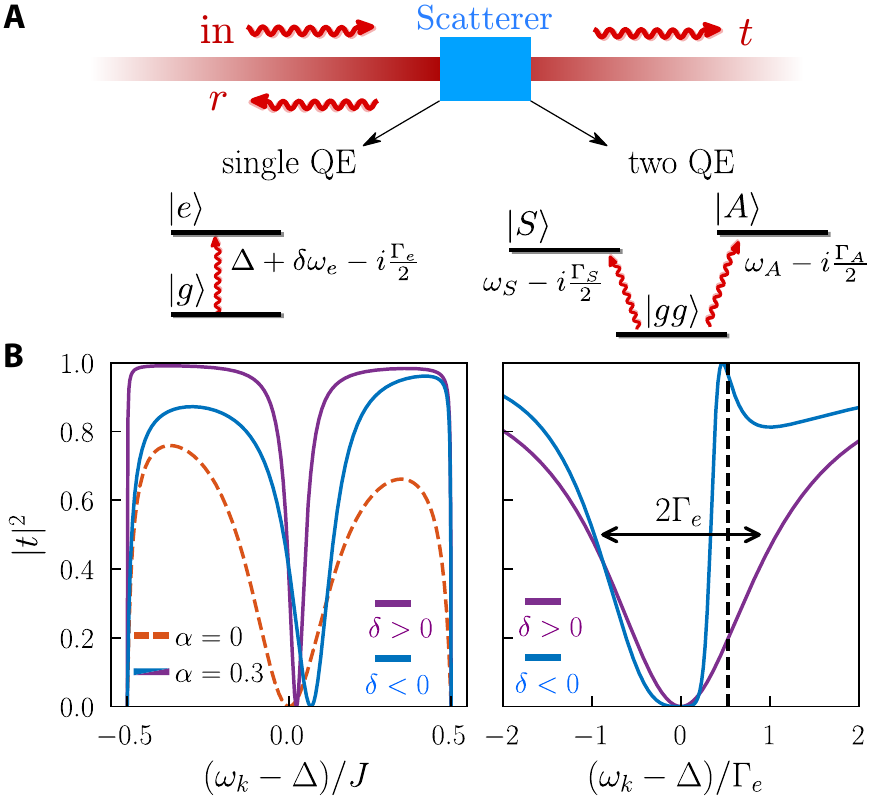}
	\caption{\textbf{Single-photon scattering}. (\textbf{A}) Pictorial
  representation of the scattering process: An incident photon impinges into a
  scatterer, part of which is reflected
  (transmitted) with probability amplitude $r$~($t$). Lower row: Relevant level
  structure for the single photon scattering for both scatterers considered: one and two
  QEs. $\ket{gg}\equiv\ket{g}_1\ket{g}_2$ denotes the common ground state, while
  $\ket{S,A}=\left(\ket{e}_1\ket{g}_2 \pm \ket{g}_1\ket{e}_2\right)/\sqrt{2}$ 
  denotes the symmetric (antisymmetric) excited state combination of the two
  QEs.
  (\textbf{B}) Transmission probability for a single emitter coupled to
  both $A$ and $B$ cavities of the same unit cell (left panel) and two emitters
  in the $AB$ configuration separated a total of $x_{12}=2$ unit cells (right
  panel). The parameters in the single emitter case are: $g=0.4J$, $\delta=\pm
  0.5$ and $\Delta=1.5J$. The dashed line corresponds to the case where the
  emitter couples to a single sublattice ($\alpha=0,1$) (does not depend on the
  sign of $\delta$). When the emitter couples to both sublattices
  ($\alpha=0.3$), the perfect-reflection resonance experiences a shift that is
  different for $\delta>0$ (purple line) or $\delta<0$ (blue line). The
  parameters for the two emitter case are: $g=0.1J$, $\delta=\pm0.5$, and
  $\Delta\simeq 1.65J$ for which the two QE are in a subradiant configuration if
  $\delta > 0$. 
  } 
	\label{fig:scattering}
\end{figure}

The scattering properties of a single photon impinging into one or several QEs
in the ground state can be obtained by solving the secular equation with
energies $H\ket{\Psi_k}=\pm\omega_k\ket{\Psi_k}$, with the $\pm$ sign depending
on the band we are probing~\cite{zhou08a}.  Here, we focus on
the study of the transmission amplitude $t$ (see scheme of
Fig.~\ref{fig:scattering}A) for two different situations: i) a single QE coupled
to both cavity $A$ and cavity $B$ in the same unit cell with coupling constants
$g\alpha$ and $g(1-\alpha)$, such that we can interpolate between the case where
the QE couples only to sublattice $A$ ($\alpha=1$) or $B$ ($\alpha=0$), ii) and
a pair of emitters in the $AB$ configuration separated $x_{12}$ unit cells.
After some algebra, we find the exact formulas for the transmission coefficients
for the two situations: 
\begin{widetext}
\begin{align}
  t_\mathrm{1QE} & =\frac{2iJ(1-\delta)\sin(k)
  \left[J(1+\delta)(\pm\omega_k-\Delta)-g^2\alpha(1-\alpha)\right]}
  {2iJ^2(1-\delta^2)(\pm\omega_k-\Delta)\sin(k)+g^2\omega_k\left[2\alpha(1-\alpha)
  (e^{-i\phi}\mp 1) \pm 1\right]}\,, \label{eq:scattering1}\\
  t_\mathrm{2QE} & =\frac{\left[2J^2(1-\delta^2)(\pm\omega_k-\Delta)\sin(k)\right]^2}
  {g^4\omega^2_ke^{i2(kx_{12}-\phi)}-\left[g^2\omega_k\pm
  i2J^2(1-\delta^2)(\pm\omega_k-\Delta)\sin(k)\right]^2}\,.
  \label{eq:scattering2}
\end{align}
\end{widetext}

In Fig.~\ref{fig:scattering}B, we plot the single-photon transmission probability 
$|t|^2$ as a function of the frequency of the incident photon for the single
(left) and two QE (right) situations. Let us now explain the different features
observed:

\emph{Single QE:} We first plot in dashed orange the results for $\alpha=0,1$,
showing well known features for this type of systems~\cite{zhou08a}, namely,
a perfect transmission dip ($|t|^2=0$) when the frequency of the incident photon
matches exactly that of the QEs. This is because the Lamb-shift induced by the
bath in this situation is $\delta\omega_e=0$. The dip has a band-width
defined by the individual decay rate $\Gamma_e$. Besides, it also shows $|t|^2=0$
at the band-edges due to the divergent decay rate at these frequencies, also
predicted for standard waveguide setups~\cite{zhou08a}. The situation becomes
more interesting for $0<\alpha<1$, since the QE energy is shifted by
$\delta\omega_e=g^2\alpha(1-\alpha)/[J(1+\delta)]$, which is different for
$\pm \delta$. This is why the dips in $|t_\mathrm{1QE}|^2$ appear at different
frequencies for $\delta=\pm 0.3$. Notice $t_\mathrm{1QE}$ is invariant under the
transformation ${\alpha\to 1-\alpha}$ (this is not true for the reflection
coefficient, which acquires a $\delta$-dependent phase shift for $\alpha=0$ but
not for $\alpha=1$).

\emph{Two QEs:} In the right panel of Fig.~\ref{fig:scattering}B we plot
$|t_\mathrm{2QE}|^2$ for two QEs coupled equally to a bath (same energy,
distance, and coupling strength), and where the only difference is the sign of
$\delta$ of the bath. The distance chosen is small such that retardation effects
do not play a significant role. The differences between $\delta>0$ and
$\delta<0$ in the $|t_\mathrm{2QE}|^2$ are even more pronounced that in
the single QE scenario since now the responses are also qualitatively different:
While the case $\delta>0$ features a single transmission dip at the QEs
frequency, for $\delta<0$ the transmission dip is followed by a window of
frequencies with perfect photon transmission, i.e, $|t_\mathrm{2QE}|^2=1$.
A convenient picture to understand this behaviour is depicted in
Fig.~\ref{fig:scattering}A, where we show that a single photon only probes the
symmetric/antisymmetric states in the single excitation subspace ($S/A$) with
the following energies (linewidths) renormalized by the bath $\omega_{S,A}=\Delta\pm
J^{AB}_{12}$ ($\Gamma_{S,A}=\Gamma_e\pm\Gamma_{12}$). For the parameters chosen
(see caption) it can be shown that for $\delta>0$ the QEs are in a perfect
super/subradiant configuration in which one of the states decouples while the
other one has $2\Gamma_e$ decay rate. Thus, at this configuration the two QEs
behave like a single two-level system with an increased linewidth. On the other hand,
when $\delta<0$ both the (anti)symmetric states are coupled to the bath, such that the
system is analogous to a V-type system where perfect transmission occurs for an
incident frequency
$\pm\omega_{k,\mathrm{EIT}}=(\omega_S\Gamma_A-\omega_A\Gamma_S)/(\Gamma_A-\Gamma_S)$
\cite{Witthaut2010} (depicted in dashed black).  

In both the single and two QE situations the different response can be
intuitively understood as the QEs couple locally to a different bath for
$\delta\gtrless 0$. However, this different response of $|t|^2$ can be thought
as an indirect way of probing topology in these systems.  

\section{Implementations~\label{sec:implementations}}

One of the attractive points of our predictions is that they can be potentially
observed in several platforms by combining tools which, in most of the cases,
have already been experimentally implemented independently. Some
candidate platforms are: 
\begin{itemize}
	\item \emph{Photonic crystals.} The photonic analogue of the SSH model has
    been implemented in several photonic
    platforms~\cite{malkova09a,jean17a,parto18a,zhao18a}, including some recent
    photonic crystal realizations~\cite{chen18a}. The latter are
    particularly interesting due to the recent advances in their integration
    with solid-state and natural atomic emitters 
    (see Refs.~\cite{lodahl15a,chang18a} and references therein).

	\item \emph{Circuit QED.} Superconducting metamaterials mimicking standard
    waveguide QED are now being routinely built and interfaced with one or many
    qubits in experiments~\cite{liu17a,mirhosseini18a}.
    The only missing piece is the periodic modulation of the couplings to obtain
    the SSH model, for which there are already proposals using circuit
    superlattices~\cite{goren18a}.  

	\item \emph{Cold-atoms.} Quantum optical phenomena can be simulated in pure
    atomic scenarios by using state-dependent optical lattices. The idea is to
    have two different trapping potentials for two atomic metastable states,
    such that one state mostly localizes, playing the role of QEs, while the
    other state propagates as a matter-wave. This proposal~\cite{devega08a} has
    been recently used~\cite{krinner18a} to explore the physics of standard
    waveguide baths. Replacing their potential by an optical superlattice
    made of two laser fields with different frequencies, one would be able to
    probe the physics of the topological SSH bath. In fact, these cold-atoms
    superlattices have already been implemented in an independent experiment to
    measure the Zak phase of the SSH model~\cite{atala13a}.  
\end{itemize}

Beyond these platforms, the bosonic analogue of the SSH model has also been
discussed in the context of metamaterials~\cite{tan14a} or
plasmonic and dielectric
nanoparticles~\cite{kruk17a,pocock18a}, where the predicted
phenomena could as well be potentially observed.  

\section{Conclusions \& Outlook~\label{sec:conclu}}

Summing up, we have presented several phenomena appearing in a topological
waveguide QED system with no analogue in other optical setups. When the
quantum emitter frequencies are tuned to the middle band-gap, we predict the
appearance of chiral photon bound states which inherit the topological
robustness of the bath. Furthermore, we also show how these bound states mediate
directional, long-range spin interactions, leading to exotic many-body phases,
e.g., double-N\'eel ordered states, which cannot be obtained to our knowledge with
other bound-state mediated interactions. Besides, we study the scattering and
super/subradiant behaviour when one or two emitters are resonant with one of
the bands, finding that transmission amplitudes can depend on the
parameter which controls the topology even though the band energy
dispersion is independent of it. 

Except for the many-body physics, the rest of the phenomena discussed in this
article, that is, the formation of chiral bound states and the peculiar
scattering properties, could also be observed in classical setups, since these
results are derived within the single-excitation regime. Given the simplicity of
the model and the variety of platforms where it can be implemented, we foresee
that our predictions can be tested in near-future experiments. 

As an outlook, we believe our work opens complementary research directions on
topological photonics, which currently focuses more on the design of exotic
light
properties~\cite{zhao18a,parto18a,jean17a,mittal16a,rechtsman16a}. For
example, the study of the emergent spin models with long-range topological
interactions is interesting on its own 
and might lead to the
discovery of novel many-body phases. Moreover, the scattering-dependent
phenomena found along the manuscript can provide alternative paths for probing
topology in photonic systems. On the fundamental level, the analytical
understanding we develop for one-dimensional systems provides a solid basis to
understand quantum optical effects in higher dimensional topological
baths~\cite{haldane88a,armitage18a}.  

\section{Acknowledgements}
\textbf{Funding:} This work has been supported by the Spanish Ministry of Economy and
Competitiveness through grants No. MAT2017-86717-P and BES-2015-071573. AGT and
JIC acknowledge the ERC Advanced Grant QENOCOBA under the EU Horizon 2020
program (grant agreement 742102).  MB, GP, and AGT  acknowledge  support  from
CSIC  Research Platform on Quantum Technologies PTI-001.
\textbf{Competing Interests:} The authors declare that they have no competing interests. 
\textbf{Data and materials availability:} All data needed to evaluate the
conclusions in the paper are present in the paper and/or the Supplementary
Materials. Additional data available from authors upon request. 
\textbf{Author contributions:} MB and AGT conceived the original idea, MB did
the analytical and numerical analysis under the supervision of AGT.  All authors
discussed and analyzed the results. MB and AGT wrote the manuscript with input
from all authors.

\newpage
 \begin{widetext}
 \widetext
\onecolumngrid
\begin{center}
	\textbf{\large Supplemental Material: Unconventional quantum optics in topological waveguide QED.}
\end{center}
\vspace{\columnsep}
\vspace{\columnsep}

\twocolumngrid
 \end{widetext}
\setcounter{equation}{0}
\setcounter{figure}{0}
\setcounter{section}{0}
\makeatletter

\makeatletter
\renewcommand{\thefigure}{S\arabic{figure}}
\renewcommand{\thetable}{S\arabic{table}}
\let\c@table\c@figure 
\renewcommand{\thesection}{S\arabic{section}}  
\renewcommand{\theequation}{S\arabic{equation}}  
\makeatother

In this Supp.\ Material, we provide more details on: i) the exact integration of
the quantum emitter (QE) dynamics using resolvent operator techniques, in
Section~\ref{secSM:inte}; ii) the study of asymptotic long-time decay, in
Section~\ref{secSM:asym}; iii) the exact integration of the two QEs dynamics, 
in Section~\ref{secSM:two}; iv) the derivation of the exact
conditions of existence of two QE bound states, in Section~\ref{secSM:twobound};
v) the effect of the edge states on the QE dynamics when they are coupled to
finite size baths, in Section~\ref{secSM:edge}; vi) a review
of bipartite one-dimensional baths and the properties of the middle
bound-states, in Section~\ref{secSM:origin}.  

\section{Integration of the dynamics~\label{secSM:inte}}

Since the global Hamiltonian $H$ conserves the number of excitations, if a QE is
initially excited with no photons in the bath, i.e.,
$\ket{\psi(0)}=\ket{e}\ket{\mathrm{vac}}$ ($\ket{\mathrm{vac}}$ denotes the
vacuum state of the lattice of bosonic modes), the wavefunction at any time has
the form: 
\begin{equation}
\ket{\psi(t)}=\left[C_e(t)\sigma_{eg}
+\sum_{j=1}^N\sum_{\alpha=a,b}C_{j,\alpha}(t)\alpha^\dagger_j\right]
\ket{g}\ket{\mathrm{vac}}\,. 
\end{equation}

\begin{figure}[tb]
  \centering
  \includegraphics{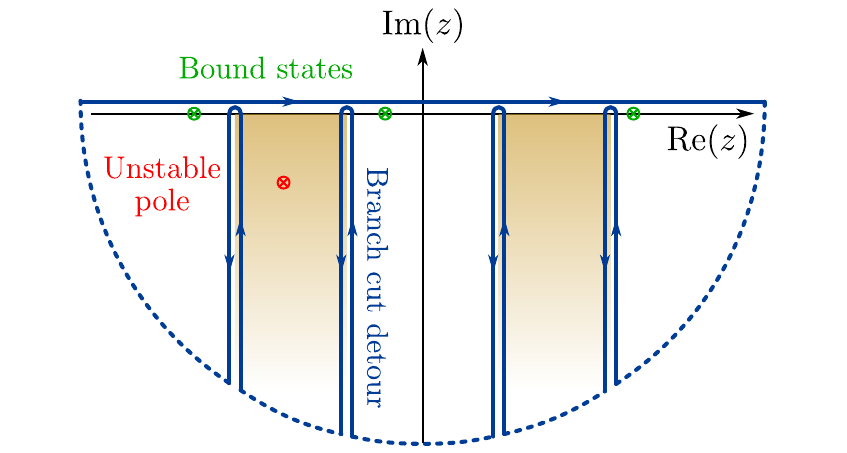}
  \caption{Schematics showing the contour of integration. At the band edges the
  path changes from the first to the second Riemann sheet of the Green function
  (shaded areas).}
  \label{fig:contour}
\end{figure}

The probability amplitude $C_e(t)$ can be 
computed~\cite{cohenbook92a,nakazato96a} as the Fourier transform of the 
Green function of the emitter $G_e(z)=[z-\Delta-\Sigma_e(z)]^{-1}$: 
\begin{equation}
C_e(t)=\frac{-1}{2\pi i}\int_{-\infty}^\infty dE\,G_e(E+i0^+)e^{-iEt}\,, 
\label{eq:integraldyn}
\end{equation}
To compute the integral in \eqref{eq:integraldyn}, we use residue integration 
closing the contour of integration in the lower half of the complex plane. 
Since the QE Green function has branch cuts in the real axis along the regions 
where the bands of the bath are defined (the continuous spectrum of $H$), it is
necessary to detour at the band edges to other Riemann sheets of the function,
see Fig.~\ref{fig:contour}.  The formula for the self-energies presented in the
main text, Eq.~\eqref{eq:singleQEselfenergy}, corresponds to the Green function
in the first Riemann sheet $G^I_e(z)$. We can analytically continue it to the
second Riemann sheet $G^{II}_e(z)$ by simply changing $\sqrt{\cdot}\rightarrow
-\sqrt{\cdot}$ in the denominator of $\Sigma_e(z)$.  

Since the imaginary part of $G^I_e(E+i0^+)$ and $G^{II}_e(E-i0^+)$ is nonzero
in the band regions, we should only take into account the real poles of $G^I_e$ 
outside the band regions ($z_\mathrm{BS}$) and the complex poles of $G^{II}_e$
with real part inside band regions ($z_\mathrm{UP}$). The residue at both the
real and complex poles can be computed as $R(z_0)=[1-\Sigma'_e(z_0)]^{-1}$,
where $\Sigma'_e(z)$ denotes the first derivative of the appropriate function
$\Sigma^I_e(z)$ or $\Sigma^{II}_e(z)$.  Finally, we should subtract the detours
taken at the branch cuts. Their contribution can be computed as 
\begin{multline}
  C_{\mathrm{BC},j}(t) = \pm\frac{1}{2\pi}\times \\ \int_0^\infty dy 
  \Big[ G^I_e(x_j-iy) - G^{II}_e(x_j-iy)\Big]e^{-i(x_j-iy)t} \,,
\end{multline}
with $x_j\in\{\pm 2J, \pm 2|\delta|J\}$. The sign has to be chosen positive if 
when going from $x_j+0^+$ to $x_j-0^+$ the integration goes from the first 
to the second Riemann sheet, and negative if it is the other way around. 

\begin{figure}
  \centering\includegraphics{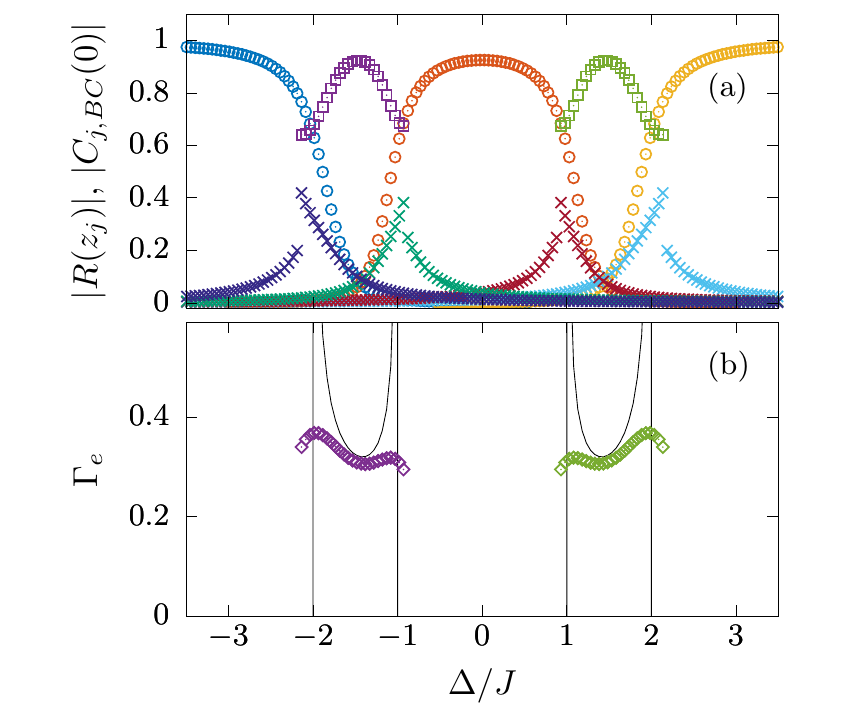}
  \caption{Non-Markovian dynamics. (a) Absolute value of the different contributions to the single QE
  dynamics at time $t=0$ as a function of the emitter detuning $\Delta$; 
  bound state residues $|R(z_\mathrm{BS})|$ (circles), unstable pole residues
  $|R(z_\mathrm{UP})|$ (squares) and branch-cut contributions 
  $|C_{\mathrm{BC},j}(0)|$ (crosses). The system parameters are 
  $\delta=0.5$ and $g=0.4J$. 
  (b) Comparison between the exact decay rate given by the imaginary part of the
  complex poles of $G_e$ (diamonds) and the approximate Markovian decay rate 
  (black lines) for the same parameters as in (a)}
  \label{fig:residues}
\end{figure}

Plotting the absolute value of the different contributions at time $t=0$, we 
can deduce the relevant physics involved in the QE dynamics, see Fig.~\ref{fig:residues}(a). 
Not surprisingly, when the emitter's transition frequency
lays in the bands of allowed bath modes it will decay emitting a photon into 
the bath. In Fig.~\ref{fig:residues}(b), we compare the actual decay rate 
with the prediction given by the Markovian approximation. On the other hand,
when it lays outside the bands, a bound state will form in which the emitter 
is mostly in the excited state and part of the photon remains trapped around 
it. This is what we observe in Fig.~\ref{fig:singleQEdynamics} in the main
text, where the long-term dynamics is dominated by the bound state at zero
energy, whose residue can be computed as $R_0=[1+g^2/(J^24|\delta|)]^{-1}$. 

\newpage
\section{Sub-exponential decay~\label{secSM:asym}}

\begin{figure}[!ht]
  \centering\includegraphics{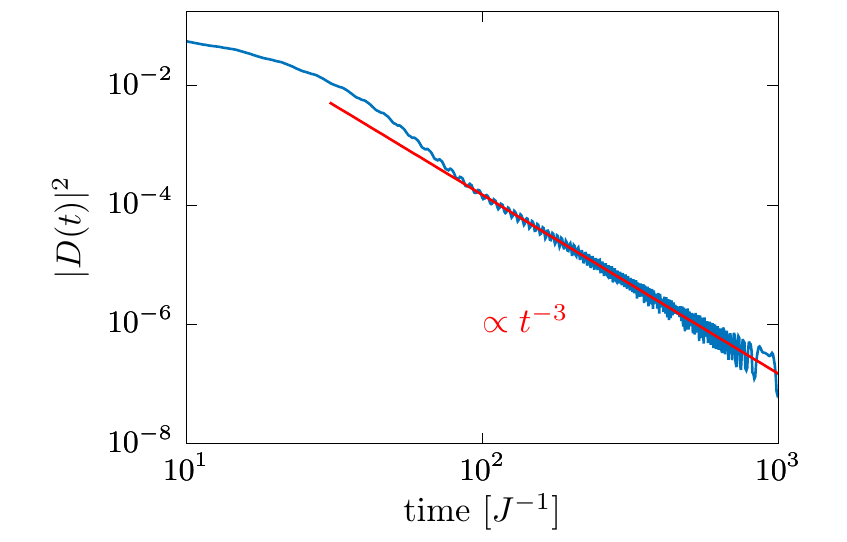}
  \caption{Decaying part of the dynamics of a single emitter with parameters 
  $\Delta=-2J$, $|\delta|=0.5$ and $g=0.2J$.}
  \label{fig:subexpdecay}
\end{figure}

Defining $D(t)\equiv C_e(t)-\sum_{z_\mathrm{BS}}R(z_\mathrm{BS})e^{iz_\mathrm{BS}t}$, at long
times we have
\begin{equation}
  \lim_{t\to\infty}D(t)\simeq\sum_j C_{\mathrm{BC},j}(t)
  =\sum_j K_j(t)e^{-ix_jt} \,,
\end{equation}
with
\begin{equation}
  K_j(t)=\frac{\pm 1}{2\pi}\int_0^\infty dy \frac{2\Sigma_e(x_j-iy)e^{-yt}}
  {(x_j-iy-\Delta)^2-\Sigma^2_e(x_j-iy)}\,. \label{eq:integrand}
\end{equation}
The long-time average of the decaying part of the dynamics can be computed as
\begin{equation}
  \overline{|D(t)|^2}\equiv\lim_{t\to\infty}\frac{1}{t}
  \int_0^t dt' |D(t')|^2=\sum_j|K_j(t)|^2 \,.
\end{equation}
If the emitter's transition frequency is close to one of the band edges, 
$\Delta\simeq x_0$, then $\overline{|D(t)|^2}\simeq|K_0(t)|^2$. In the long-time
limit, we can expand the integrand in power series around $y=0$,
\begin{multline} 
  K_0(t)=\\\frac{\pm1}{2\pi}\int_0^\infty dy\left[
    \frac{4}{g^2}\sqrt{\frac{i(2-x^2_0+2\delta^2)}{x_0}} + \mathcal{O}(y)\right]
  y^{1/2}e^{-yt}\\
  \simeq \frac{\pm1}{\sqrt{\pi g^2}}\sqrt{\frac{i(2-x^2_0+2\delta^2)}{x_0}}
  t^{-3/2}+\mathcal{O}(t^{-5/2})\,.
\end{multline}
Therefore, to leading order $\overline{|D(t)|^2}\sim t^{-3}$. In 
Fig.~\ref{fig:subexpdecay} it is shown an example of this algebraic decay 
when $\Delta$ is placed at the lower band edge of the bath's spectrum. 

\section{Two QE dynamics in the non-Markovian regime~\label{secSM:two}}

The dynamics of two emitters are not much harder to analyze than that of a
single emitter. It can be shown that the symmetric and antisymmetric combinations 
$\sigma^\dagger_\pm=\big[\sigma^{1}_{eg}\pm\sigma^{2}_{eg}\big] /\sqrt{2}$ 
couple to orthogonal bath modes~\cite{gonzaleztudela17b}. 
Therefore, the two-emitter problem can be split in two independent single-emitter 
problems. The Green functions associated to the probability amplitudes 
to find the 1st or the 2nd emitter in the excited state $C_{1,2}(t)$ can be 
obtained form the Green functions associated to the symmetric/antisymmetric 
combination of excitations as ${G_{1,2}(z)=[G_+(z)\pm G_-(z)]/2}$, with
${G_\pm(z)=[z-\Delta-\Sigma_\pm(z)]^{-1}}$.

Rewriting the interaction Hamiltonian in the bath's eigenmode basis,
substituting $\sigma^{m}_{eg}$ in terms of $\sigma^\dagger_\pm$, and pairing
the terms that go with opposite momentum, we obtain for the case where the two
QEs are on the sublattice $A$
\begin{widetext}
\begin{gather}
  H^{AA}_I = \frac{g}{\sqrt{N}}\sum_{k>0}\sum_{\beta=\pm}\sqrt{1+\beta\cos(kx_{12})}
  (\tilde{u}_{k,\beta} + \tilde{l}_{k,\beta})\sigma^\dagger_\beta + \mathrm{H.c.} 
  \,, \label{eq:HIAA_ini}\\
  \tilde{u}_{k,\pm} = \frac{1}{2\sqrt{1\pm\cos(kx_{12})}}
  \left[\left(e^{i(kx_1+\phi)}\pm e^{i(kx_2+\phi)} \right)u_k
  + \left(e^{-i(kx_1+\phi)}\pm e^{-i(kx_2+\phi)} \right)u_{-k}\right] \,, \\
  \tilde{l}_{k,\pm} = \frac{1}{2\sqrt{1\pm\cos(kx_{12})}}
  \left[\left(e^{i(kx_1+\phi)}\pm e^{i(kx_2+\phi)} \right)l_k
  + \left(e^{-i(kx_1+\phi)}\pm e^{-i(kx_2+\phi)} \right)l_{-k}\right] \,,
  \label{eq:HIAA_end}
\end{gather}
Here, $x_n$ refers to the unit cell where the $n$'th QE is located, and 
$x_{12}=x_2-x_1$ is the signed distance between the two QEs. For the case where
the two QEs are on a different sublattice
\begin{gather}
  H^{AB}_I = \frac{g}{\sqrt{N}}\sum_{k>0}\sum_{\beta=\pm}
  \left[ \sqrt{1+\beta\cos(kx_{12}-\phi)}
  \, \tilde{u}_{k,\beta}\sigma^\dagger_\beta + \sqrt{1-\beta\cos(kx_{12}-\phi)}\,
  \tilde{l}_{k,\beta}\sigma^\dagger_\beta \right] + \mathrm{H.c.} \,, 
  \label{eq:HIAB_ini}\\
  \tilde{u}_{k,\pm} = \frac{1}{2\sqrt{1\pm\cos(kx_{12}-\phi)}}
  \left[\left(e^{i(kx_1+\phi)}\pm e^{ikx_2} \right)u_k
  + \left(e^{-i(kx_1+\phi)}\pm e^{-ikx_2} \right)u_{-k}\right] \,, \\
  \tilde{l}_{k,\pm} = \frac{1}{2\sqrt{1\mp\cos(kx_{12}-\phi)}}
  \left[\left(e^{i(kx_1+\phi)}\mp e^{ikx_2} \right)l_k
  + \left(e^{-i(kx_1+\phi)}\mp e^{-ikx_2} \right)l_{-k}\right] \,.
  \label{eq:HIAB_end}
\end{gather}
\end{widetext}
The prefactors in the definition of $\tilde{u}_{k,\pm}$ and $\tilde{l}_{k,\pm}$
come from normalization. Importantly, these modes are orthogonal, they
satisfy 
\begin{equation}
  [\tilde{u}_{k,\alpha},\tilde{u}^\dagger_{k',\alpha'}]=
  [\tilde{l}_{k,\alpha},\tilde{l}^\dagger_{k',\alpha'}]=
  \delta_{kk'}\delta_{\alpha\alpha'} \,.
\end{equation}
Since $\omega(k)=\omega(-k)$, we have that the bath Hamiltonian is also diagonal
in this new basis. The two other configurations can be analyzed analogously. 

From these expressions for the interaction Hamiltonian, it is possible to obtain
the self-energy for the symmetric/antisymmetric states of the two QE.
As it turns out, they have a very simple form: 
$\Sigma^{\alpha\beta}_\pm=\Sigma_e\pm\Sigma^{\alpha\beta}_{12}$, with 
$\Sigma^{\alpha\beta}_{12}$:
\begin{align}
  \Sigma^{AA/BB}_{mn}(z;x_{mn})=\frac{g^2}{N}\sum_k\frac{ze^{ikx_{mn}}}{z^2-\omega^2(k)}
\,, \label{eq:selfeAABB}\\
\Sigma^{AB}_{mn}(z;x_{mn})=\frac{g^2}{N}\sum_k\frac{\omega(k)e^{i[kx_{mn}-\phi(k)]}}
  {z^2-\omega^2(k)} \,, \label{eq:selfeAB}
\end{align}
where $x_{mn}=x_n-x_m$. It can be shown that 
\begin{multline}
\Sigma^{BA}_{mn}(z; \delta, x_{mn}) = \Sigma^{AB}_{nm}(z; \delta, -x_{mn}) \\ 
  = \Sigma^{AB}_{mn}(z; -\delta, x_{mn} - 1)\,.
\end{multline}

\section{Existence conditions of two QE bound states~\label{secSM:twobound}}

We can integrate the dynamics in the same way as we did for the single QE 
case, but there are some subtleties particular to the two QE case. First, the 
cancellation of divergences of $\Sigma_e$ and $\Sigma^{\alpha\beta}_{12}$ at some of 
the band edges results in critical transition frequencies above (or below) 
which some bound states cease to exist. For example, in the symmetric 
subspace we have that the lower bound state ($E_\mathrm{BS}<-2J$) always 
exists, while the upper bound state ($E_\mathrm{BS}>2J$) exists only for 
$\Delta>\Delta^\mathrm{out}_c$, 
\begin{equation}
\Delta^\mathrm{out}_c=2J-\frac{g^2(2x_{12}+1-\delta)}{2J(1-\delta^2)}\,.
\end{equation}
For the middle bound state there are two possibilities: either the divergence 
vanishes at $-2|\delta|J$, in which case the bound state will exist for 
$\Delta>\Delta^\mathrm{mid}_c$, or the divergence vanishes at $2|\delta|J$,
then the middle bound state exists for $\Delta<\Delta^\mathrm{mid}_c$. In 
both cases $\Delta^\mathrm{mid}_c$ takes the same form
\begin{equation}
  \Delta^\mathrm{mid}_c=(-1)^{x_{12}}\left\{2\delta J
  + \frac{g^2[(2x_{12}+1)\delta-1]}{2J(1-\delta^2)}\right\} \,,
\end{equation}

\begin{figure}
	\centering\includegraphics{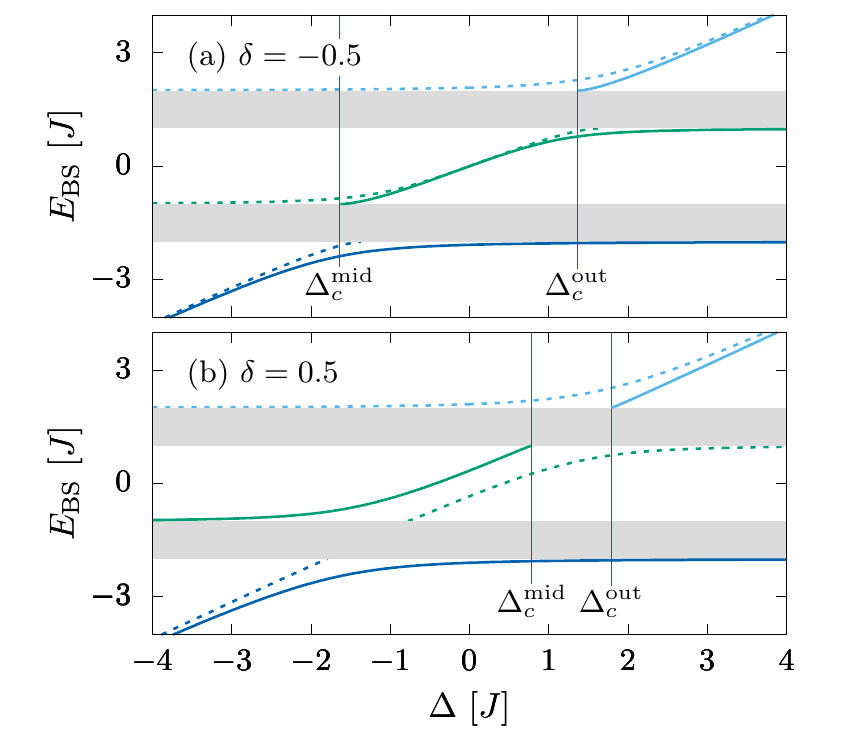}
	\caption{Bound states for the symmetric (continuous) and the antisymmetric
		(dashed) subspaces as a function of the QEs transition frequency in the
		topological (a) and the trivial (b) cases; $g=0.8J$ and $x_{12}=0$.
    The shaded areas mark the bath's band regions, where no bound states can be found.}
	\label{fig:2QEboundstates}
\end{figure}

The situation in the antisymmetric subspace can be readily understood realizing
that $\Re\Sigma^{AB}_-(z)=-\Re\Sigma^{AB}_+(-z)$, which implies that if 
$z=E_\mathrm{BS}$ is a solution of the pole equation for $\Sigma^{\alpha\beta}_+$ with a
particular value of $\Delta$, then $z=-E_\mathrm{BS}$ is a solution of the pole
equation for $\Sigma^{\alpha\beta}_-$ with the opposite value of the emitter transition
frequency. Fig.~\ref{fig:2QEboundstates} summarizes at a glance the different
possibilities and the dependence on the bath's topology.

\section{Finite-bath dynamics~\label{secSM:edge}}

It is well known that a finite bath with open boundary conditions in the
topologically non-trivial phase ($\delta<0$) supports a pair of edge states
$\ket{\mathrm{ES}_\pm}$, with opposite energies
$H_B\ket{\mathrm{ES}_\pm}=\pm\epsilon\ket{\mathrm{ES}_\pm}$, given by
$\epsilon\simeq J(1-\delta)e^{-N/\lambda}$. These states are exponentially localized
at the edges of the bath with the same localization length $\lambda$ as the BSs at
zero energy mentioned in the main text. So far, our calculations have been done
in baths large enough such that the contributions of the topological edge-states
could be neglected. In this section, we consider the effect they can have in
systems with moderate sizes.  

\begin{figure*}
	\centering\includegraphics{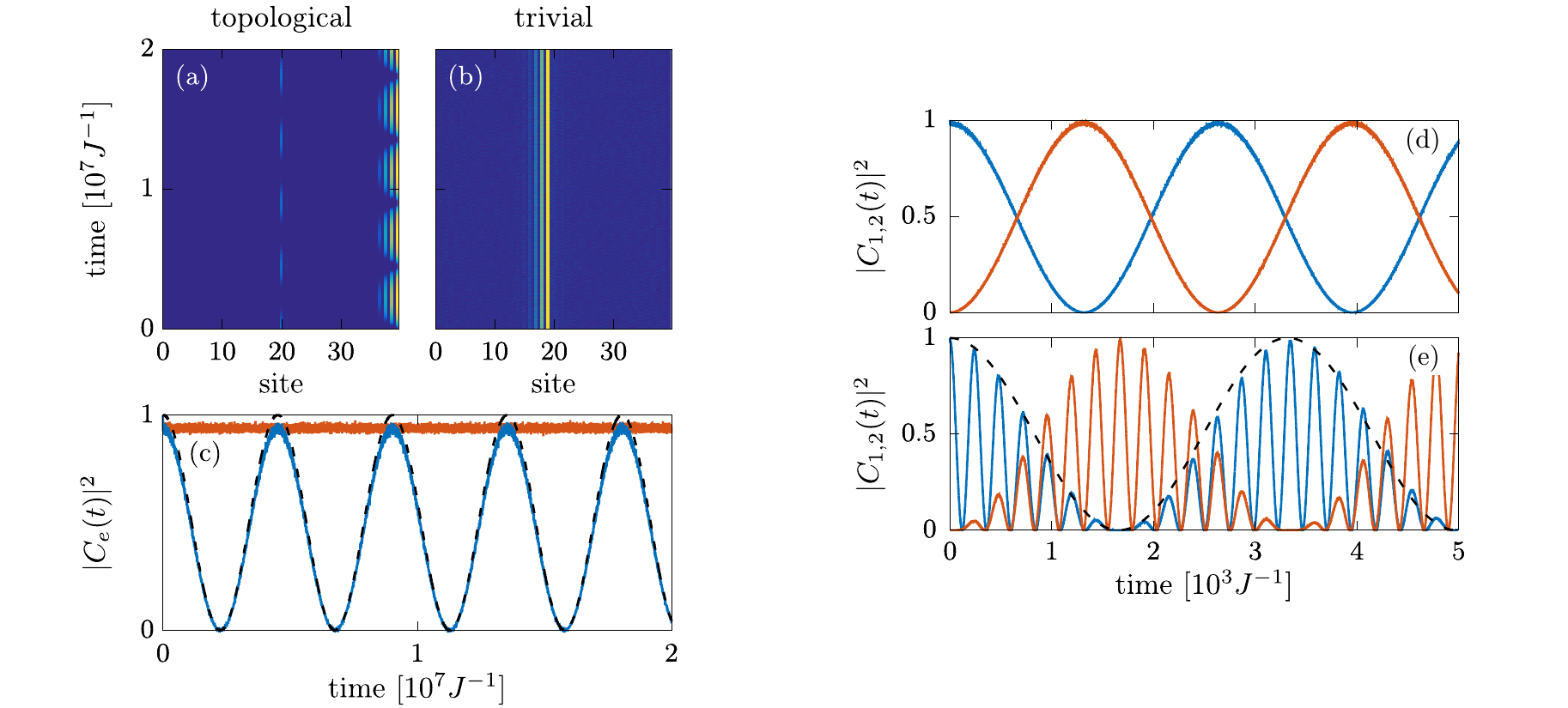}
  \caption{Finite-size effects. (Left panel) Bath dynamics in the topological, $\delta=-0.3$, (a) and trivial,
  $\delta=0.3$ (b) regimes, for a lattice with $N=40$ unit cells and open boundary conditions.
  A single QE with $\Delta=0$ is coupled with strength $g=0.2J$ to the middle of the bath.
  The color shows the probability to find the photon in each site of the
  lattice. Brighter colors correspond to a higher probability. We have used a
  different logarithmic scale in each case for clarity. Below (c), it is shown
  the probability to find the emitter in the excited state for both the
  topological (blue) and trivial (red) cases. The dashed black line is a cosine
  function with frequency $2\tilde{\omega}_0$, as obtained by a more precise
  treatment using Green functions. (Right panel) Dynamics for two QEs coupled to the $A/B$ 
  lattices respectively, placed symmetrically around the middle of the bath 
  ($\Delta=0$) with parameters $N=10$, $g=0.1J$, $x_{12}=3$, and $\delta=0.3$ 
  (d), $\delta=-0.3$ (e). The dashed black line is a cosine with a frequency 
  obtained from the exact treatment with Green functions.} 
  \label{fig:singleQE_open_dynamics}
\end{figure*}

In Fig.~\ref{fig:singleQE_open_dynamics}(a--c) we compare the dynamics of an
initially excited QE coupled to a finite bath ($N=40$) in the topologically
non-trivial and trivial phases with the same $|\delta|=0.3$. The induced
dynamics is very different: while most of the QE excitation remains localized
around the QE for a topologically trivial bath, in the non-trivial case the QE
exchanges non-locally the excitation with one of the edges of the bath. This
emergent dynamics can be captured by a simple effective Hamiltonian considering
only the excited state of the QE and the two edge states (with the QE in the
ground state):
\begin{equation}
H_\mathrm{eff}=\begin{pmatrix}
\Delta & \tilde{g}_+ & \tilde{g}_- \\
\tilde{g}_+ & \epsilon & 0 \\
\tilde{g}_- & 0 & -\epsilon
\end{pmatrix} \,,
\label{eq:Heff1}
\end{equation}
written here in the basis $\{\ket{e}\ket{\mathrm{vac}},
\ket{g}\ket{\mathrm{ES}_+}, \ket{g}\ket{\mathrm{ES}_-}\}$. The coupling
constants are
$\tilde{g}_\pm=g\bra{\mathrm{ES}_\pm}c^\dagger_{x_1}\ket{\mathrm{vac}}$
($c^\dagger_{x_1}$ is equal to $a^\dagger_{x_1}$ or $b^\dagger_{x_1}$ depending
on the sublattice to which the emitter is coupled) and satisfy
$|\tilde{g}_-|=|\tilde{g}_+|\equiv\tilde{g}$. Exactly when $\Delta=0$, the QE
couples more strongly to the edge states. In that case, the excited-state
probability amplitude can be computed as 
\begin{equation}
C_e(t)\simeq\frac{\epsilon^2+2\tilde{g}^2\cos(\omega_0t)} 
{\epsilon^2+2\tilde{g}^2} \,,
\end{equation}
with $\omega_0=\sqrt{\epsilon^2-2\tilde{g}^2}$. 
Note that a (anti)symmetric  superposition of the edge states corresponds to an
exponentially localized state in one of the ends of the chain. Due to this, the
photon oscillates between the QE and the edge whose ending mode is in the
sublattice to which the QE is coupled [see Fig.~\ref{fig:singleQE_open_dynamics}(a)]. 
The oscillation frequency given by
the effective model overestimates the actual frequency, which can be calculated
exactly using the resolvent operator formalism. We do so by extending the bath,
adding the two edge states, which are orthogonal to all other bath modes. The
emitter Green function becomes now 
\begin{equation}
  G_e=\frac{z^2-\epsilon^2}{(z-\Delta-\Sigma_e)(z^2-\epsilon^2)
  -2\tilde{g}^2z}\,.
\end{equation}
The long-term dynamics is given just by the real poles of this modified Green
function. In particular, for $\Delta=0$ the denominator is and odd function
with three real roots around the middle of the band gap: $z=0$ and
$z=\pm \tilde{\omega}_0$. It can be shown that the largest contribution
to the dynamics comes from these real poles, such that $C_e(t)\simeq
R_0+2R_+\cos(\tilde{\omega}_0 t)$, where $R_0$ denotes the
residue at the pole $z=0$, and $R_+=R_-$ is the residue at the poles 
$z=\pm \tilde{\omega}_0$.


In Fig.~\ref{fig:singleQE_open_dynamics}(d, e), we show the QE population
dynamics when two QEs are coupled to the $A/B$ lattices symmetrically with
respect to the middle of the chain, and for two different situations, i.e., with
fixed $|\delta|=0.3$ but different sign. As in the individual behaviour, the
collective dynamics is very different depending on the topological nature of the
bath. In the topologically trivial bath, the BS mediates perfect coherent
transfer of excitations between the two QEs [see
Fig.~\ref{fig:singleQE_open_dynamics}(d)].  In the topologically non-trivial
bath, however, the edge states become largely populated since they are
quasi-resonant with the QE oscillation, leading to additional oscillatory
behaviour. Interestingly, perfect coherent transfer is still possible at certain
times [see Fig.~\ref{fig:singleQE_open_dynamics}(e)], even though the induced
dipolar coupling is zero. This dynamics can again be captured by a simple
effective Hamiltonian, which written in the basis 
$\{\ket{e_1}\ket{g_2}\ket{\mathrm{vac}}, 
\ket{g_1}\ket{e_2}\ket{\mathrm{vac}},
\ket{g_1}\ket{g_2}\ket{\mathrm{ES}_+}, \ket{g_1}\ket{g_2}\ket{\mathrm{ES}_-}\}$ reads
\begin{equation}
H_\mathrm{eff}=\begin{pmatrix}
\Delta & J^{AB}_{12} & \tilde{g} & \tilde{g} \\
J^{AB}_{12} & \Delta & \tilde{g} & -\tilde{g} \\
\tilde{g} & \tilde{g} & \epsilon & 0 \\
\tilde{g} & -\tilde{g} & 0 & -\epsilon 
\end{pmatrix}\,,
\end{equation} 
using the definitions of the edge-states and the coupling constants
$\tilde{g}_\pm$ for each QE that we use in the single QE dynamics.  Solving this
Hamiltonian with $\Delta=0$, and assuming  $\epsilon\ll \tilde{g}$, the excited
state occupation probability of the 1st (2nd) emitter can be well approximated
by: 
\begin{align}
C_{1}(t) &\approx  \cos(\epsilon
  t/2)\cos\big(\sqrt{2}\tilde{g}t\big)\,,\\
C_{2}(t) &\approx  \sin(\epsilon
  t/2)\cos\big(\sqrt{2}\tilde{g}t\big)\,.
\end{align}
which captures qualitatively the double oscillatory behaviour of
Fig.~\ref{fig:singleQE_open_dynamics}(e). In order to quantitatively capture the
frequencies of the transfer exactly, one can use resolvent operator techniques,
which yields the dashed black line of Fig.~\ref{fig:singleQE_open_dynamics}(e).  
In this case, the extended Green functions are given by 
\begin{equation}
  G_\pm=\frac{z\pm\epsilon}{(z-\Delta-\Sigma^{\alpha\beta}_\pm)(z\pm\epsilon)-2\tilde{g}^2}\,.
\end{equation}
For $\Delta=0$, the real poles of $G_+$ around the middle of the band gap,
$z_\pm$, are the same as those of $G_-$ with opposite sign. The residues are the
same in both cases, therefore $C_\pm(t)\simeq R_+e^{\pm iz_+t}+R_-e^{\pm
iz_-t}$. Since $C_{1,2}(t)=[C_+(t)\pm C_-(t)]/2$, the relevant frequencies are
$\omega_\pm=||z_+|\pm|z_-||$.

It should be noted, however, that for really small systems or situations in
which the emitters are placed close to the edges, the results given by these
modified Green functions will not be accurate, as they use the thermodynamic
self-energies $\Sigma^{\alpha\beta}_\pm$, which are obtained for infinite systems. \\

\section{Middle bound states in one-dimensional baths~\label{secSM:origin}}

\begin{table*}
  \centering
  \begin{tabular}{M{5cm}M{4cm}M{4cm}M{1cm}M{1cm}M{1cm}}
    \textbf{Model} & \textbf{Quantized Zak Phase} & \textbf{Bulk-Boundary Correspondence} 
                      & \textbf{A} & \textbf{B} & \textbf{C} \\ \hline\hline
    Coupled cavity array with staggered energies &  No & Not applicable & No  & No  & No   \\ \hline
    SSH                                          & Yes & Yes            & Yes & Yes & Yes  \\ \hline
    SSH with staggered energies                  & No  & Not applicable & Yes & Yes & No   \\ \hline
    SSH with next-nearest neighbours             & Yes & No             & Yes & No  & Yes* \\ \hline
  \end{tabular}
  \label{tab:sum}
  \caption{Topological properties of several one-dimensional baths, and their
    corresponding bound state features A--C (see text for discussion) when an
    emitter couples to them.}
\end{table*}
	
The central part of the manuscript analyzes the properties and consequences of a
peculiar bound state which appears in the middle band-gap when an emitter with
energy $\Delta$ couples to the bath. In particular, this bound state has the
following properties: A) it is chiral, in the sense that it localizes
preferentially in one side of the emitter; B) when $\Delta=0$, the bound state
has its energy in the middle of the gap, $E_{\mathrm{BS}}=0$, being fully
directional and with no amplitude in the sublattice to which the QE couples; C)
it inherits the topological robustness to disorder from the bath.  

To make evident that the photonic SSH model is the simplest one-dimensional model
where all these properties are satisfied, and connect it with the topological
features of the bath, let us consider a general bipartite bath Hamiltonian
defined by: 
\begin{equation}
\tilde{H}_B(k) = \left( \begin{array}{cc}
G_A(k) & F(k) \\
F^*(k) & G_B(k)
\end{array} \right)\,.
\label{eq:HB}
\end{equation}

Depending on the functions $G_{A/B}(k),F(k)$, this Hamiltonian covers a plethora
of relevant one-dimensional models with a middle band-gap where an extra
bound-state appears. In Table~\ref{tab:sum} we review the topological properties
of several of these models, and whether middle bound states show the features
A--C discussed above: 
\begin{enumerate}
	
	\item When $G_{A/B}(k)= \omega_a\pm\delta\omega$, $F(k)=-J$, we have the
    simple coupled-cavity array model with staggered energies. The staggered
    energies break the symmetry between the $A$/$B$ sublattices, opening a middle
    band-gap. In this case, an extra bound-state can be found in the middle of
    the bands but with none of the features A--C.  

  \item Setting 
    \begin{align*}
      & G_{A/B}(k) =\omega_a\,, \\ 
      & F(k) =-J[(1+\delta)+(1-\delta)e^{-ik}]\,,
    \end{align*}
    we recover the SSH model
    whose topological and bound state properties were already discussed in the
    main text.  
  
  \item One can combine the SSH model with staggered energies, i.e.,
    \begin{align*}
      & G_{A/B}(k) =\omega_a \pm \delta\omega \,, \\ 
      & F(k) =-J[(1+\delta)+(1-\delta)e^{-ik}]\,,
    \end{align*}
    which still shows a
    middle band-gap but chiral symmetry is broken. The staggered energies allow
    one to go from one band-configuration to the other without closing the gap,
    such that the Zak phase is not quantized anymore. The middle bound states
    can be chiral, but they do not get the robustness to disorder since the
    model is topologically trivial.  
  
  \item Finally, we consider the SSH model adding next-nearest neighbour
    hoppings, i.e.,  
    \begin{align*}
      & G_{A/B}(k)=\omega_a-2J_2\cos(2k)\,, \\
      & F(k)=-J\left[(1+\delta)+(1-\delta)e^{-ik}\right] \,.
    \end{align*}
    This model does not preserve chiral symmetry, but it does preserve 
    spatial inversion symmetry. The later still leads
    to a quantized Zak phase, but bulk-boundary correspondence is not guaranteed
    any more (the number of edge states is not linked to the topological
    invariant)~\cite{Bea2018}. Even though the bound states are chiral, they are
    not fully directional, and they are only robust to disorder of very
    restricted type (one which respects spatial inversion symmetry).
\end{enumerate}



\newpage
\bibliography{photonicSSH,Sci,books}

\end{document}